\documentclass[sigconf]{acmart}
\renewcommand\footnotetextcopyrightpermission[1]{} 
\def\BibTeX{{\rm B\kern-.05em{\sc i\kern-.025em b}\kern-.08emT\kern-.1667em\lower.7ex\hbox{E}\kern-.125emX}}

\usepackage[utf8]{inputenc}
\usepackage{pdfpages}
\usepackage{todonotes}
\usepackage{xspace}
\usepackage{listings}
\usepackage{pgfplots}
\usepackage{pgfplotstable}
\usepackage{subcaption}
\usepackage{amsmath}
\usepackage{bm}      
\usepackage{xcolor}  

\usepackage{paralist}

\usepackage[T1]{fontenc}
\usepackage{lmodern}

\definecolor{cbgreen}{RGB}{27,158,119}
\definecolor{cbred}{RGB}{217,95,2}
\definecolor{cbblue}{RGB}{117,112,179}

\hypersetup{draft}

\newcommand{\octo}{Octo-Tiger\xspace}
\newcommand{\octotiger}{Octo-Tiger\xspace}

\newcommand{\futurization}{Futurization\xspace}
\newcommand{\amt}{Asynchronous Many Task\xspace}
\newcommand{\cxxppm}{C++ Parallel Programming Model\xspace}
\newcommand{\future}{\lstinline{future}\xspace}
\newcommand{\futures}{\lstinline{future}s\xspace}

\newcommand{\channels}{\lstinline{channel}s\xspace}
\newcommand{\regtd}{\textsuperscript{\textregistered}}
\newcommand{\tradem}{\textsuperscript{\texttrademark}}

\begin{document}
\setdefaultleftmargin{9pt}{}{}{}{}{}

%
\title{From Piz Daint to the Stars: Simulation of Stellar Mergers using High-Level Abstractions}

%

\author{Gregor Dai\ss}
\affiliation{\institution{IPVS, University of Stuttgart}}
\email{Gregor.Daiss@ipvs.uni-stuttgart.de}

\author{Parsa Amini}
\affiliation{\institution{CCT, Louisiana State University}}
\email{parsa@cct.lsu.edu}
\authornotemark[1]

\author{John Biddiscombe}
\affiliation{\institution{Swiss National Supercomputing Centre}}
\email{biddisco@cscs.ch}
\authornotemark[1]

\author{Patrick Diehl}
\affiliation{\institution{CCT, Louisiana State University}}
\email{pdiehl@cct.lsu.edu}
\orcid{0000-0003-3922-8419}
\authornotemark[1]

\author{Juhan Frank}
\affiliation{\institution{Louisiana State University}}
\email{frank@phys.lsu.edu}

\author{Kevin Huck}
\affiliation{\institution{University of Oregon}}
\email{khuck@cs.uoregon.edu}

\author{Hartmut Kaiser}
\affiliation{\institution{CCT, Louisiana State University}}
\email{hkaiser@cct.lsu.edu}
\authornote{The STE$||$AR Group, {\tt stellar-group.org}}

\author{Dominic Marcello}
\affiliation{\institution{CCT, Louisiana State University}}
\email{dmarcello@phys.lsu.edu}
\authornotemark[1]

\author{David Pfander}
\affiliation{\institution{IPVS, University of Stuttgart}}
\email{David.Pfander@ipvs.uni-stuttgart.de}

\author{Dirk Pfl\"{u}ger}
\affiliation{\institution{IPVS, University of Stuttgart}}
\email{dirk.pflueger@ipvs.uni-stuttgart.de}

%
\renewcommand{\shortauthors}{Dai\ss, et al.}

\begin{abstract}
We study the simulation of stellar mergers, which requires complex
simulations with high computational demands. We have developed
Octo-Tiger, a finite volume grid-based hydrodynamics simulation code
with Adaptive Mesh Refinement which is unique in conserving both
linear and angular momentum to machine precision. To face the
challenge of increasingly complex, diverse, and heterogeneous HPC
systems, Octo-Tiger relies on high-level programming abstractions. 

We use HPX with
its futurization capabilities to ensure scalability both between nodes
and within, and present first results replacing MPI with libfabric
achieving up to a 2.8x speedup.
We extend Octo-Tiger to heterogeneous GPU-accelerated supercomputers,
demonstrating node-level performance and portability. We show
scalability up to full system runs on Piz Daint. For the
scenario's maximum resolution, the compute-critical parts (hydrodynamics and
gravity) achieve 68.1\% parallel efficiency at 2048 nodes.
\end{abstract}



%
%
%


\begin{teaserfigure}
   \includegraphics[width=\textwidth]{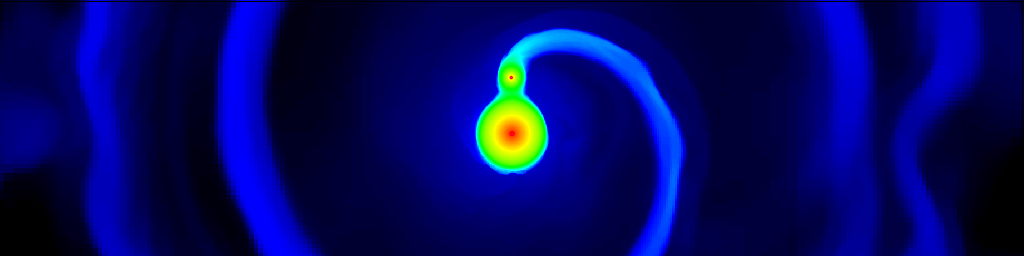}
   \caption{The \octo\ model of V1309 Scorpii 20 orbits after the simulation begins. V1309 Scorpii is a contact binary
       that merged into a single star in 2008 in a process known as a luminous red nova. It was the first star
       to provide conclusive evidence that contact binary systems end their evolution in a stellar merger~\cite{tylenda_2011}, see also Section~\ref{sec:scenario}.}
   \label{fig:teaser}
 \end{teaserfigure}

%
\maketitle

\section{Introduction}
\label{sec:introduction}

Astrophysical simulations are among the classical drivers for exascale
computing. They require multiple scales of physics and cover
vast scales in space and time. Even the next generation of
high-performance computing (HPC) systems will be insufficient to
solve more than a fraction of the many conceivable scenarios.

However, new HPC systems come not only  with ever larger processor counts, but increasingly complex, diverse, and
heterogeneous hardware. Evolving manycore architectures and GPUs are combined with
multicore systems. This raises challenges especially for large-scale
HPC simulation codes
and requires going beyond traditional programming models. High-level
abstractions are required to ensure that codes are portable and can be
run on current HPC systems without the need to rewrite large portions of
the code.

We consider the simulation of stellar phenomena based on the
simulation framework \octo. In particular, we study the simulation of
time-evolving stellar mergers (Fig.~\ref{fig:teaser}). The study of binary
star evolution from the onset of mass transfer to merger
can provide fundamental insight into the underlying physics. In 2008,
this phenomenon was observed with photometric data, when the contact
binary V1309 Scorpii merged to form a luminous red novae
\cite{tylenda_2011}. The vision of our work is to model this event with simulations on
HPC systems.
Comparing the results of our simulations with the observations will enable us to validate
the model and to improve our understanding of the physical processes involved.

\octo\ is an HPC application and relies on
high-level abstractions, in particular, HPX and Vc. While HPX
provides scheduling and scalability, both between nodes and within, Vc
ensures portable vectorization across processor-based platforms.
To make use of GPUs we use HPX's CUDA integration in this work.

Previous work has demonstrated scalability on Cori, a Cray
XC40 system installed at the National Energy Research Scientific
Computing Center (NERSC) ~\cite{heller_gb}.
However, the underlying node-level
performance was rather low, and they were only able to simulate for few
time steps. Consequently, they had started to study
node-level performance, achieving 408 GFLOPS on the 64 cores of the
Intel Knights Landing manycore processor~\cite{Pfander18accelerating}. Using the same
high-level abstractions as on multicore systems, this led to a speedup
of 2 compared to a 24-core Intel Skylake-SP platform.


In this work, we make use of the same CPU level abstraction library Vc ~\cite{k2015}
for SIMD vector parallelism as in the previous study, but extend 
\octo to support GPU-based HPC machines.
We show how the critical node-level bottleneck, the fast multipole method (FMM) kernels, can be mapped to GPUs.
Our approach utilizes GPUs as co-processors, running up to 128 FMM kernels on each one simultaneously.
This was implemented using CUDA streams and uses HPX's futurization approach for lock-free, low-overhead scheduling.
We demonstrate the performance-portability of \octo for a set of GPU and processor-based HPC nodes.

To scale more efficiently to thousands of nodes, 
we have integrated a new libfabric communication backend 
into HPX where it can be used transparently by \octo ~-- the first large scientific
application to use the new network layer.
The libfabric implementation extensively uses one-sided communication to reduce the overhead compared to a standard two-sided MPI-based backend.
To demonstrate both our node-level GPU capabilities as well as our improved scalability with libfabric, we show results for full-scale runs on Piz Daint running the real-world stellar merger scenario of V1309 Scorpii for a few time-steps.
Piz Daint is a Cray XC40/XC50 equipped with NVIDIA's P100 GPUs at the Swiss National Supercomputing Centre (CSCS). For our full system runs we used up to 5400 out of 5704 nodes.
This is the first time an HPX application was run on a full system of a GPU-accelerated supercomputer.

In Sec.~\ref{sec:related} we briefly discuss related approaches.
We describe the stellar scenario in more detail in Sec.~\ref{sec:scenario}, 
the important parts of the overall software framework and the
high-level abstractions they provide in Sec.~\ref{sec:framework}.
In turn, Sec.~\ref{sec:imp} shows the main contributions of this work, describing both the new libfabric parcelport and the way we utilize GPUs to accelerate the execution of critical kernels. 
In Sec.~\ref{sec:results:node-level}, we present our node-level performance 
results for NVIDIA GPUs, Intel Xeons and an Intel Xeon Phi platform.
Section~\ref{sec:results} describes our scaling results, showing that we are able to scale with both an MPI communication backend and a libfabric communication backend of HPX.
We show that the use of libfabric strongly improves performance at scale.
\vspace{-0.5em}
\section{Related work}
\label{sec:related}


There are several studies that investigate the structure of mass loss in V1309
Scorpii through computer simulation.
One approach to modeling this system is smoothed-particle
hydrodynamics (SPH).
Notable SPH applications include StarSmasher~\cite{lombardi_2016,nandez_2008}
(a fork of StarCrash~\cite{faber2010starcrash}) and an unpublished code
developed by a collaboration of researchers from Princeton University, Columbia
University, and Osaka University~\cite{pejcha2015cool,pejcha_2017}.
An alternative approach is to use the finite volume
method to simulate mass transfer.
Examples of such applications include
Athena~\cite{stone2008athena,athena_repo} and its rewrite named
Athena++~\cite{athena++,macleod_2018,macleod_2018_2}. Lastly,
Enzo~\cite{bryan2014enzo} is a project that implements finite
volume hydrodynamics along with a collisionless N-body module that can be
used to simulate binary systems where one component is taken to be a point
mass. With the exception of SPH codes using direct summation for gravity,
\octo is unique among three-dimensional self-gravitating hydrodynamics codes
in that it simultaneously conserves both linear and angular momentum to 
machine precision. SPH codes using direct summation for gravity are limited 
to only a few thousand particles, making \octo the 
better choice for high resolution simulations.

Adaptive multithreading systems such as HPX expose
concurrency by using user-level threads.
Some other notable solutions that take such an approach are
Uintah~\cite{germain2000uintah}, Chapel~\cite{chamberlain2007parallel},
Charm++~\cite{kale1993charm++}, Kokkos~\cite{CarterEdwards20143202},
Legion~\cite{bauer2012legion}, and PaRSEC~\cite{bosilca2013parsec}. 
Note that we only refer to distributed memory capable solutions, 
since we focus here on large distributed simulations. Different
task-based parallel programming models, e.g.\ Cilk Plus, OpenMP, Intel TBB, Qthreads, StarPU, GASPI, Chapel, Charm++, and HPX, are compared in~\cite{thoman2018taxonomy}. Our requirements (distributed, task-based, asynchronous) are met by few, out of which
HPX has the highest technology readiness level according to this review. It
is furthermore the only one with a future-proof C++ standard conforming
API and allows us to support the libfabric networking library without
changing application code. For more details, see Sec.~\ref{subsec:hpx}.

There are several particle-based FMM implementations utilizing
task-based programming available. The approach described
in~\cite{ltaief2014data} uses the Quark runtime
environment~\cite{yarkhan2011quark}, 
the implementation in \cite{agullo2016task,agullo2016taskreport} uses
StarPu~\cite{augonnet2011starpu}, whilst \cite{choi2014cpu} uses
OpenMP~\cite{Dagum:1998:OIA:615255.615542}, and
\cite{zhang2014asynchronous} compares Cilk~\cite{blumofe1996cilk}, HPX-5,
and OpenMP tasks~\cite{openmptr}. 
Our choice of HPX for the task-based runtime system is 
motivated by the same findings as the above 
mentioned review and the need to implement specialized
kernels for energy conservation that require coupling
between different parts of the solver.

While conservation of linear momentum to machine precision is possible with
existing FMM implementations, \octo employs a novel extension to
the FMM that also ensures conservation of angular momentum to machine
precision (see Sec.~\ref{section:octotiger}). 
Another extension requires a solution for the time-derivative of 
the gravitational field to ensure conservation of total energy.
The coupling of the gravitational derivative with the
hydrodynamics solver in turn requires the use of a volume-based
FMM code (making integration of a particle-based code very challenging); 
HPX's futurization technique makes this coupling straightforward
while maintaining efficient hardware utilization.
Additionally, the planned addition of radiation transport and 
other solvers in the future can also take advantage of the unique 
futurization properties of HPX. 
None of the other available task-based FFM 
implementations examined, such as PVFMM, ExaFMM-alpha, minifmm, or DASHMM 
met the requirements for integration into \octo. 
The best fitting and scalable of the alternative candidates would 
be the volume-based PVFMM code; however, it uses Chebyshev polynomials 
of higher degree, which results in a significantly higher flops/cell 
rate than our implementation which assumes locally homogeneous densities. 
In \octo, it would be possible to
use the surrounding leaf cells to compute higher order multipole
moments at the leaf cell level, resulting in a higher computational
density and better GPU performance (as for PVFMM).
However, it is not clear how to ensure the conservation
of all momenta for polynomials of higher degree.
For the reasons cited here, we have developed new FMM kernels, 
compatible with HPX for this work.



\vspace{-0.5em}
\section{Scenario: Stellar Mergers}
\label{sec:scenario}
In September 2008, the contact binary, V1309 Scorpii, merged to form a
luminous red novae (LRN)~\cite{tylenda_2011}. The Optical Gravitational Lensing
Experiment (OGLE) observed this binary prior to its merger, and six years of data show 
its period decreasing. When the
merger occurred, the system increased in brightness by a factor of
about $5000$. Mason et al.~\cite{mason_2010} observed the outburst
spectroscopically, confirming it as a LRN. This was the first observed
stellar merger of a contact binary with photometric data available prior to
its merger.

Possible progenitor systems for V1309 Scorpii, consisting initially of
zero-age main sequence stars with unequal masses in a relatively
narrow range, were proposed by Stepien in~\cite{stepien_2011}.  As the
heavier of the two stars first begins to expand into a red giant, it
transfers mass to its lower mass companion, forming a common
envelope. The binary's orbit shrinks due to friction, and the mass
transfer slows down as the companion becomes the heavier of the two
stars but continues to grow at the expense of the first
star. Eventually this star also expands, with both stars now touching
each other forming a contact binary.
Stepien et. al. 
sampled the space of physically possible initial masses, finding that
initial primary masses of between $1.1 M_\odot$ and $1.3 M_\odot$ and 
initial secondary masses between $0.5 M_\odot$ and $0.9 M_\odot$ produced
results consistent with observations prior to merger. The evolution described above
results in an approximately $1.52 - 1.54 M_\odot$ primary and a $0.16 - 0.17 M_\odot$
secondary with helium cores and Sun-like atmospheres.  
It is
theorized that the merger itself was due to the Darwin
instability. When the total spin angular momentum of a binary system
exceeds one third of its orbital angular momentum, the system can no
longer maintain tidal synchronization. This results in a rapid tidal
disruption and merger.
\octo\ uses its Self-Consistent Field module \cite{hachisu,even_and_tohline} 
to produce an initial
model for V1309 to simulate this last phase of dynamical evolution. 
The stars are tidally synchronized, and the stars have a common
atmosphere. The system parameters are chosen such that the spin angular momentum
just barely exceeds one third of the orbital angular momentum.
\octo\ begins the simulation just as the Darwin instability sets in (Fig.~\ref{fig:teaser}).



\vspace{-0.5em}
\section{Software Framework}
\label{sec:framework}

\subsection{HPX}
\label{subsec:hpx}

We have developed the \octotiger application framework~\citep{octotiger_git} in
  ISO C++11 using HPX~\citep{heller:2012,Heller:2013:UHL:2530268.2530269,hpx_pgas_2014,espm2_2015,heller:hpc_2016,hpx_git}.
HPX is a C++ standard library for distributed and parallel programming built on
  top of an \amt (AMT) runtime system. Such AMT runtimes may provide a means
  for helping programming models to fully exploit available parallelism on
  complex emerging HPC architectures. The HPX methodology described here includes
 the following essential components:

\begin{compactitem}
\item An ISO C++ standard conforming API that enables wait-free asynchronous parallel programming, including \futures, \channels, and other primitives for asynchronous execution.
\item An Active Global Address Space (AGAS) that supports load balancing via object migration and enables exposing a uniform API for local and remote execution.
\item An active-message networking layer that enables running functions close to the objects they operate on. This also implicitly overlaps computation and communication.
\item A work-stealing lightweight task scheduler that enables finer-grained parallelization and synchronization and automatic load balancing across all local compute resources.
\item APEX, an in-situ profiling and adaptive tuning framework.
\end{compactitem}
\smallskip 

The design features of HPX allow application developers to naturally
  use key parallelization and optimization techniques, such as overlapping communication and computation,
  decentralizing control flow, oversubscribing execution resources, and
  sending work to data instead of data to work. As a result \octotiger achieves 
  exceptionally high system utilization and exposes very good weak- and strong scaling
  behaviour.
HPX exposes an asynchronous, standards conforming programming model enabling \futurization, 
  with which developers can express complex dataflow execution
  trees that generate billions of HPX tasks that are scheduled to execute only when their
  dependencies are satisfied~\cite{heller_gb}. Also, \futurization\ 
  enables automatic parallelization and load-balancing to emerge.
Additionally, HPX provides a performance counter and adaptive tuning framework that
  allows users to access performance data, such as core utilization, task
  overheads, and network throughput; these diagnostic tools
  were instrumental in scaling \octo to the full machine.

This paper demonstrates the viability of the HPX programming model at scale using
\octotiger, a portable and standards conforming application. \octotiger fully embraces
the \cxxppm, including additional constructs that are  incrementally
being adopted into the ISO C++ Standard.
The programming model views the entire supercomputer as a single C++ abstract
  machine.
A set of tasks operates on a set of C++ objects distributed across the system.
These objects interact via asynchronous function calls; a function call to an
  object on a remote node is relayed as an active message to that node.
A powerful and composable primitive, the \future object represents
  and manages asynchronous execution and dataflow.

A crucial property of this model is the semantic and syntactic equivalence of
  local and remote operations.
This provides a unified approach to intra- and inter-node
  parallelism based on proven generic algorithms and data structures
  available in today's ISO C++ Standard.
The programming model is intuitive and enables performance portability
across a broad spectrum of increasingly diverse HPC hardware.

\vspace{-0.5em}
\subsection{\octo}
\label{section:octotiger}

\octo\ simulates the evolution of mass density, momentum, and
energy of interacting binary stellar systems from the start of mass
transfer to merger. It also evolves five passive scalars.
It is a three-dimensional finite-volume code with Newtonian
gravity that simulates binary star systems as self-gravitating compressible inviscid fluids.
To simulate these fluids we need three core components:
(1) a hydrodynamics solver, (2) a gravity solver that calculates the gravitational field produced by the fluid distribution, and (3) a solver to generate an initial configuration of the star system.


The passive scalars, expressed in units of mass density,
are evolved using the same 
continuity equation that describes the evolution of the mass density.
They do not influence the flow itself, but are rather used to track
various fluid fractions as the system evolves. In the case of V1309,
these scalars are initialized to the mass density of the 
accretor core, the accretor envelope, the donor core, the donor envelope,
and the common atmosphere between the two stars. The passive scalars
are useful in post-processing. For instance, to compute the temperature
we require the mass and energy densities as well as the number density.
The latter is not evolved in the simulation, but can be computed
from the passive scalars assuming a composition for each fraction (e.g.
helium for both cores, and a solar composition for the remaining fractions).

The balance of angular momentum plays an important role in the orbital evolution 
of binary systems. Three-dimensional astrophysical fluid codes with self-gravity 
do not typically conserve angular momentum. The magnitude of this violation is 
dependent on the particular problem and resolution. Previous works have found 
relative violations as high as $10^{-3}$ per orbit 
\cite{Motl_2002, marcello_and_tohline, 2011apj...737...89d}. This error, accumulated 
over several dozen orbits, becomes significant enough to influence the fate of 
the system. \octo conserves both linear and angular momenta to machine precision. 
In the fluid solver, this is accomplished using a technique described by 
\cite{despres2015}, while the gravity solver uses our own extension to the FMM.

\octo's main datastructure is a rotating Cartesian grid with adaptive mesh refinement (AMR). It is
based on an adaptive octree structure. Each node is an $N^3$ sub-grid (with $N=8$ for all runs in this paper) containing the evolved variables, and can be further refined into eight
child nodes. Each octree node is implemented as an HPX component.
These octree nodes are distributed onto the compute nodes using a space filling curve. For further information about implementational details we refer to~\cite{Pfander18accelerating} and ~\cite{octotiger_apcs_2016}.

The first solver that operates on this octree is a finite volume hydrodynamics solver. \octo uses 
the central advection scheme of \cite{2000jcoph.160..241k}. The piece-wise parabolic method (PPM) 
\cite{1984JCoPh..54..174C} is used to compute the thermodynamic variables at cell faces. A method 
detailed by \cite{marcello_and_tohline} is used to conserve total energy in its interaction with 
the gravitational field. This technique involves applying the advection scheme to the sum of gas 
kinetic, internal, and potential energies, resulting in conservation of the total energy. 
Numerical precision of internal energy densities can suffer greatly in high mach flows, where the 
kinetic energy dwarfs the gas internal energy. We use the dual-energy formalism of 
\cite{bryan2014enzo} to overcome this issue: We evolve both the gas total 
energy as well as the entropy. The internal energy is then computed from one or the other 
depending on the mach number (entropy for high mach flows and total gas energy for low mach ones). The angular momentum technique described by \cite{despres2015} is applied to the 
PPM reconstruction. It ads a degree of freedom to the reconstruction of velocities on 
cell faces by allowing for the addition of a spatially constant angular velocity component to the
linear velocities. This component is determined by evolving three additional variables corresponding
to the spin angular momentum for a given cell.


The gravitational field solver is based on the FMM.
\octo\ is unique in conserving both linear and angular
momentum simultaneously and at scale using modifications to the original FMM algorithm \cite{octotiger_fmm,octotiger_apcs_2016}.

Finally, we assemble the initial scenario using the Self-Consistent Field technique alongside the FMM solver.
\octo\ can produce initial models for binary systems that are in contact, semi-detached, or detached~\cite{octotiger_apcs_2016}.
Calculated only once, the computational demands of this solver will be negligible for full-size runs.

We used a test suite of four verification tests, recommended by Tasker et al.\ \cite{tasker} for self-gravitating astrophysical codes, to verify the correctness of our results.
The first two are purely hydrodynamic tests: the Sod shock tube and the Sedov-Taylor blast wave.
Both have analytical solutions which we can use for comparisons.
The third and fourth test are a globular star cluster in equilibrium
and one in motion. In each case, the equilibrium structure should be
retained.
Because \octo is intended to simulate individual stars self-consistently, we have substituted a single star in equilibrium at rest for the third test and a single star in
equilibrium in motion for the fourth test.

\subsection{The FMM hotspot}
\label{sec:fmm}
The most compute-intensive task is the calculation of the gravitational field using the FMM, since this has to be done for each of the fluid-solver time-steps.
Note that our FMM variant differs from approaches such as the implementation used in~\cite{YOKOTA2013445}.
While being distributed and GPU-capable, their FMM is operating upon particles. Our FMM variant operates on the grid cells directly since each grid cell has a density value which determines its mass, and thus its gravitational influence on other cells.
We further differ from other (cell-based) FMM variants used for computing gravitational fields by conserving not only linear momentum, but also angular momentum, down to machine precision using the changes outlined in \cite{octotiger_fmm}.
Due to its computational intensity, we will take a closer look at the FMM and its kernels in this section. 

The FMM algorithm consists of three steps. First, it computes the multipole moments and the center-of-masses of the individual cells. This information is then used to calculate Taylor-series expansions coefficients in the second and third steps. These coefficients can in turn be used to approximate the gravitational potential in a cell, which can then be used by the hydrodynamics solver \cite{octotiger_apcs_2016}.

The first of the three FMM steps requires a bottom up traversal of the
octree datastructure. 
The fluid density of the cells of the highest level is the starting point. The multipole moments of every other cell are then calculated using the multipole moments of its child cells. We can additionally compute the center of mass for each refined cell.
While this step includes a tree-traversal, it is not very compute intensive.

In the second FMM step (same-level), we use the multipole moments and the
center-of-masses to calculate how much the gravity in each cell is
influenced by its neighboring cells on the same octree level. How many
cells are considered as ``neighboring'' is determined by the
so-called opening criteria~\cite{octotiger_apcs_2016}. However, their number is constant on each level. 
The result of these interactions is a Taylor series expansion of interactions.
This is the most compute-intensive part.

In the third FMM step, the gravitational influence of cells outside of the opening criteria is computed, and the octree is traversed top-down. The respective Taylor series expansion of the parent node is passed to the child nodes and accumulated.

In the first and third step we calculate interactions between either child nodes and their respective parents or vice-versa. Since a refined node only has 8 children, the number of these interactions is limited. In the second step, the number of same-level interactions per cell that need to be calculated is much higher. For our choice of parameters, each cell interacts with 1074 of its close neighbors, assuming they exist.


The second FMM step (same-level interactions) is by far the most compute-intense part. Originally, it required about 70\% of the total scenario runtime and was thus the core focus of previous optimizations.
Originally, lookup of close neighbor cells was performed using an
interaction list, and data was stored in an array-of-struct format. In
order to improve cache-efficiency and vector-unit usage, we changed it
to a stencil-based approach and are now utilizing a struct-of-arrays
datastructure. Compared to the old interaction-list approach, this led
to a speedup of the total application runtime between 1.90 and 2.22 on AVX512
CPUs and between 1.23 and 1.35 on AVX2
CPUs~\cite{master_thesis_2018_daiss}. Furthermore, we achieved
node-level scaling as well as performance portability between
different CPU architectures through the usage of Vc
\cite{Pfander18accelerating, master_thesis_2018_daiss}. After these
optimizations, the FMM required only about 40\% (depending on the
hardware) of the total scenario runtime with its compute kernels
reaching a significant fraction of peak on multiple platforms as we will
demonstrate in Sect.~\ref{sec:results:node-level}.

Due to the presence of AMR, there are four different cases of same-level
interactions: 1) multipole-monopole interactions between cells of a refined octree node 
(multipoles) and cells of a non-refined octree node
(monopoles); 2) multipole-multipole interactions; 3) 
monopole-monopole interactions; and 4) monopole-multipole interactions.
This yields four kernels per
octree-node. Their input data are the current node's sub-grid as well
as all sub-grids of all neighboring nodes as a halo (ghost layer).
The kernels then compute all interactions of a certain type and add
the result to the Taylor coefficients of the respective cells in the
sub-grid. 
We were able to combine the multipole-multipole and the
multipole-monopole kernels into a single kernel, yielding three
compute kernels in our implementation.

As the monopole-multipole kernel consumes only about 2\% of the
total runtime, we ignore it in the following. The remaining
two compute kernels, 1)/2) and 3), are the central hotspots of the
application. Each kernel launch applies a 1074 element stencil for each
cell of the octree's sub-grid. As we
have $N^3=512$ cells
per sub-grid, this results in $549\,888$ interactions per kernel
launch. Depending on the interaction type, each of those interactions
requires a different number of floating point operations to be
executed. For monopole-monopole interactions we execute 12 floating
point operations per interaction, and for multipole-multipole/monopole
interaction 455 floating point operations. More information about the 
kernels can be found in \cite{Pfander18accelerating}; however, the
number of floating operations per monopole interaction differs
slightly there as we combined the two monopole-X kernels there.

\section{Improving \octo using high-level abstractions}
\label{sec:imp}
Running an irregular, adaptive application like \octo on a heterogeneous supercomputer like Piz Daint presents challenges: The pockets of parallelism contained in each octree node must be run efficiently on the GPU, despite the relatively small number of cells in each sub-grid. 
The GPU implementation should not degrade parallel efficiency through overheads such as work aggregation, CPU/GPU synchronization, or blocked CPU threads. 
Furthermore, we expect the implementation to behave as before, with the exception of faster GPU execution of tasks.

In this section, we first present our implementation and integration 
of FMM GPU kernels into the task flow using HPX CUDA futures as a high-level abstraction.
We then introduce the libfabric parcelport and show how 
this new communication layer improves scalability of \octo by taking advantage
of HPX's communications abstractions.

\vspace{-0.2em}
\subsection{Asynchronous Many Tasks with GPUs}
\label{sec:gpu}
 As our FMM implementation is stencil-based and uses a struct-of-arrays datastructure, the FMM kernels introduced in Section \ref{sec:fmm} are very amenable to GPU 
execution.
  Each kernel executes a 1074 element stencil on the 512 cells of the 8x8x8 sub-grid of an octree node, calculating the gravitational interactions of each cell with its 1074 neighbors.
  We parallelize over the cells of the sub-grid, launching kernels with 8 blocks, each containing 64 CUDA threads which execute the whole stencil for each cell.
  The stencil-based computation of the interactions between two cells is done the same way as on the CPU. In fact, since we use Vc datatypes for vectorization on the CPU, we can simply instance the same function template (that computes the interaction between two cells) with scalar datatypes and call it within the GPU kernel.
  GPU-specific optimizations are done in a wrapper around this cell-to-cell method and the loop over the stencil elements. This wrapper includes the usual CUDA optimizations such as shared and constant memory usage.

  Thus far we have used standard CUDA to create rather normal kernels for the FMM implementation. However, these kernels alone suffer from two major issues:
  As it stands, the execution of a GPU kernel would block the CPU
  thread launching it, no other task would be scheduled or executed whilst it runs.
As \octo relies on having thousands of tasks available simultaneously
for scalability, this presents a problem.
  The second issue is obvious when looking at the size of the workgroups and the number of blocks for each GPU kernel launch mentioned above. The GPU kernels do not expose enough parallelism to fully utilize a GPU such as the NVIDIA P100 using only small workgroups and 8 blocks per kernel.
  To solve these two issues, we provide an HPX abstraction for CUDA streams.

For any CUDA stream event we create an HPX future that becomes ready once 
operations in the stream (up to the point of the event/future's creation) are finished.
Internally, this is created using a CUDA callback function that sets the future ready \cite{heller:hpc_2016}.
This seemingly simple construct allows us to fully integrate CUDA kernels within
the HPX runtime, as it provides a synchronization point for 
the CUDA stream that is compatible with the HPX scheduler. 
It yields multiple immediate advantages:
 \begin{compactitem}
   \item Seamless and automatic execution of kernels and overlapping of CPU/GPU tasks;
   \item overlapping of computation and communication as some HPX
     tasks are related to the communication with other compute nodes; and
   \item CPU/GPU data synchronization - completed GPU kernels triggering the scheduler, signal access to buffers that can be used/copied.
 \end{compactitem}
 Furthermore, the integration is mostly non-invasive since
a CUDA kernel invocation now equates to a function call returning a future. 
The rest of the kernel implementation and the (asynchronous) buffer handling uses the normal CUDA API, thus the GPU kernels themselves can still be hand-optimized.
Nonetheless, this integration alone does not solve the second issue: The kernels are too fine-grained to fully utilize the GPUs.
Conventional approaches to solve this include work aggregation and execution models where CUDA kernels can call other kernels and coalesce execution.

 Unfortunately, work aggregation schemes, as described in \cite{orr2014fine}, do not fit our task-based approach. 
Individual kernels should finish as soon as possible in order to trigger dependent ones, such as communication with other nodes 
or the third FMM step; delays in launching these may lead to a degradation of parallel efficiency.
 Recursively calling other GPU kernels as in \cite{wang2016dynamic} poses a similar problem as we would traverse the octree on the GPU, making communication calls more difficult. Furthermore, we would like to run code on the appropriate device; tree traversals on the CPU, and processing of the octree kernels on the GPU.

Here, however, we can exploit the fact that the execution of GPU kernels is just another task to the HPX runtime system: 
We launch a multitude of different GPU kernels concurrently 
on different streams with each CPU thread handling multiple CUDA streams, 
and thus multiple GPU kernels concurrently.
Normally, this would present problems for CPU/GPU synchronization 
as GPU results are needed for other CPU tasks. But the 
continuation passing style of program execution in HPX, 
chaining dependent tasks onto futures, makes this trivial.
When a GPU kernel output (or data transfer) that has not yet finished
is needed for a task, the runtime assigns different work to the CPU
and schedules the dependent tasks when the GPU future becomes ready.
When the number of concurrent GPU tasks running matches the total number of available CUDA streams (usually 128 per GPU), new kernels are instead executed as CPU tasks 
until a CUDA stream becomes empty again.

In summary, the octree is traversed on the CPU, with tasks spawned
asynchronously for kernels on the GPU or CPU returning futures for each. 
Any tasks that require results from previous ones are attached as continuations to the futures. The CPU is continuously supplied with 
new work (including communication tasks) as futures complete.
Since all CPU threads may participate in traversal and
steal work from each other, we keep the GPU busy by nature of
the sheer number of concurrent GPU kernels submitted.

\octo is the first application to use HPX CUDA futures. It is in fact an ideal fit for this kind of GPU integration:
Parallelization is possible only within individual timesteps of the application, 
and a production run simulation will require tens of thousands of them, 
making it is essential to maximize parallel efficiency 
(as well as proper GPU usage), particularly as each timestep might run for
a fraction of a second on the whole machine overall. The fine-grained approach of GPU usage presented here fits these challenges perfectly.

 In Section \ref{sec:results:general} we show how this model performs. We run a real-world scenario for a few timesteps to both show that we achieve a significant fraction of GPU peak performance during the execution of the FMM, as well as scalability on the whole Piz Daint machine, each of the 5400 compute nodes using a NVIDIA P100 GPU.
 Thus, \octo also serves as a proof as concept, showing that large, tree-based applications containing pockets of parallelism can efficiently run fine-grained parallelism tasks on the GPU without compromising scalability with HPX.

\vspace{-0.2em}
\subsection{Active messages and libfabric parcelport}
\label{sec:libfrabric}
The programming model of HPX does not rely on the user matching network sends and receives explicitly
as one would do with MPI. Instead, active messages are used to transfer data and
trigger a function on a remote node; we refer to the triggering of remote functions with
bound arguments as actions and the messages containing the serialized data and remote
function as parcels~\cite{ac:2017}.
A halo exchange, for example, written using MPI involves a receive operation posted
on one node and a matching send on another. With non-blocking MPI operations,
the user may check for readiness of the received data at a convenient place
in the code and then act appropriately. With blocking ones,
the user must wait for the received data and can only continue as soon as it arrives.

With HPX, the same halo exchange may be accomplished by creating a \future for some
data on the receiving end, and having the sending end trigger an action that sets the
\future \textit{ready} with the contents of the parcel data.
Since \futures in HPX are the basic synchronization primitive for work, the user
may attach a continuation to the receive data to start the next
calculation that depends on it.
The user does not therefore have to perform any test for readiness of the
received data: When it arrives, the runtime will set the future and schedule whatever work
depends upon it automatically. This combines the convenience of both a blocking receive
to trigger work, with an asynchronous receive that allows the runtime to continue
whilst waiting.

The asynchronous send/receive abstraction in HPX has been extended
with the concept of a \textit{channel} that the receiving end may fetch futures from
(for $N$ timesteps ahead if desired) and the sending end may push data into as it is generated.
Channels are set up by the user similar to MPI communicators;
however, the handles to channels are managed by AGAS (Sect. \ref{subsec:hpx}).
Even when a grid cell is migrated from one node to another during operation,
the runtime manages the updated destination address transparently, allowing the user 
code to send data to the relocated grid with minimal disruption.
These abstractions greatly simplify user level code and 
allow performance improvements in the runtime to be propagated seamlessly
to all places that use them.

The default messaging layer in HPX is built on top of the asynchronous two-sided
MPI API and uses Isend/Irecv within the parcel encoding and decoding steps of
action transmission and execution. HPX is designed from the ground up to be
multi-threaded, avoid locking/waiting, and instead suspend tasks and execute others
as soon as any blocking activity takes place.
Although MPI supports multi-threaded applications, it has its own internal
progress/scheduling management and locking mechanisms that
interfere with the smooth running of the HPX runtime.
The scheduling in MPI is in turn built upon the network provider's asynchronous
completion queue handling and multi-threaded support which may also use OS level
locks that suspend threads (and thus impede HPX progress).

The HPX parcel format is more complex than a simple MPI message, but the overheads
of packing data can be kept to a minimum \cite{ac:2017} by using remote memory access (RMA) for transfers.
All user/packed data buffers larger than the eager message size threshold are encoded
as pointers and exchanged between nodes using one-sided RMA put/get operations.
Switching HPX to use the one-sided MPI RMA API 
is no solution as this involves memory registration/pinning that is passed through
to the provider level API, causing additional (unwanted) 
synchronization between user code,
MPI code, and the underlying network/fabric driver.
Bypassing MPI and using the network API directly to improve performance was seen
as a way of decreasing latency, improving memory management,
simplifying the parcelport code, and better integrating
the multi-threaded runtime with the communications layer.
Libfabric was chosen as it has an ideal API that is supported on many platforms, including
Cray machines via the GNI provider~\cite{pritchard2016gni}.

The purely asynchronous API of libfabric blends seamlessly with the asynchronous internals
of HPX. Any task scheduling thread may poll for completions in libfabric and set futures
to received data without any intervening layer. A one-to-one mapping of
completion events to ready futures is possible for some actions, and
dependencies for those \futures can be immediately scheduled for execution.
We expose pinned memory buffers for RMA to libfabric via allocators in the HPX
runtime, so that internal data copying between user buffers (halos for example)
and the network is minimized.
When dealing with GPUs capable of multi TFlop performance, even delays of the order of
microseconds in receiving data and subsequent task launches translates to a significant loss
of compute capability.
Note that with the HPX API it is trivial to reserve cores for thread pools dedicated
to background processing of the network separate from normal task execution to
further improve performance, but this has not yet been attempted with the \octotiger code.

Our libfabric parcelport uses only a small subset of the libfabric API but delivers
very high performance as we demonstrate in Sect.~\ref{sec:results}.
It should be stressed that the improvements we see in throughput are more a result
of switching from two to one-sided communication, rather than
abandoning MPI. Similar gains could probably be made using the MPI RMA
API, but this would require a much more complex implementation.

It is a significant contribution of this work that  
we have demonstrated that an application may benefit from significant 
performance improvements in the runtime without changing a single line 
of the application code. This has been achieved utilizing 
abstractions for task management, scheduling, distribution, and messaging.
It is generally true of any library that improvements
in performance will produce corresponding improvements in code using
it. But switching a large codebase to one-sided or asynchronous
messaging is usually a major operation that involves redesigns
of significant portions to handle synchronization between
previously isolated (or sequential) sections.
The unified futurized and asynchronous API of HPX provides 
a unique opportunity to take advantage of improvements at all
levels of parallelism throughout a code as all tasks
are naturally overlapped. And network bandwidth and latency 
improvements reduce waiting not only for remote data, 
but the effects of improved scheduling of all messages 
(synchronization of remote tasks as as well as direct data 
movement) directly impacts and improves on-node scheduling 
and thus benefits all tasks.


\section{Results}
\label{sec:results:general}
The initial model of our V1309 simulation includes a $1.54 M_\odot$ primary and a $0.17 M_\odot$ secondary.
Each have helium cores and solar composition envelopes, and there is a common envelope surrounding both 
stars. The simulation domain is a cubic grid with edges $1.02 \times 10^3 R_\odot$ long. This
is about 160 times larger than the initial orbital separation, providing space for any mass ejected from the
system. The sub-grids are
$8 \times 8 \times 8$ grid cells. The centers of mass
of the components are $6.37 R_\odot$ apart. The grid is rotating about the z-axis with a
period of $1.42$ days, corresponding to the initial period of the binary. For the level $14$ run, both stars
are refined down to $12$ levels, with the core of the accretor and donor refined to $13$ and $14$
levels respectively. The $15$, $16$, and $17$ level runs are successively refined one more level in each
refinement regime. At the finest level, each grid cell is $7.80 \times 10^{-3} R_\odot$ in each dimension
for level $14$, down to $9.750 \times  10^{-4} R_\odot$ for
level $17$. Although available compute time allowed us only to
simulate a few time-steps for this work, this is exactly the production
scenario we aim for.
For all obtained results, the software dependencies in Table~\ref{tab:octo_dependencies} were used to build \octo (d6ad085) on the various platforms.

\begin{table}[tb]
\begin{tabular}{lc|lc}
\toprule
HPX        & 45f3d80 & Vc         & 1.4.1  \\
Boost      & 1.68.0  & hwloc      & 2.0.3  \\
GCC        & 7.3.0   & tcmalloc/gperftools   & 2.7  \\
Cray-MPICH & 7.7.2   & HDF5       & 1.10.4 \\
Silo       & 4.10.2  & libfabric  & 1.7.0  \\
CUDA       & 9.2     & cmake & 3.12.0 \\
\bottomrule
\end{tabular}

  \caption{Software dependencies of \octo (d6ad085).}
  \label{tab:octo_dependencies}
 \vspace{-0.75cm}
\end{table}

\subsection{FMM Node-Level Performance}
\label{sec:results:node-level}
\begin{table*}[t]
  \centering
\begin{tabular}{l|c|c|c|c|c}
\toprule
Utilized Hardware                                       & Execution    & Total scenario & \multicolumn{3}{c}{FMM}                   \\
                                                        &              & runtime        & runtime & GFLOP/s      & fraction of peak \\
\midrule
Intel\regtd Xeon\tradem E5-2660 v3 , 2.4 GHz, 10 Cores  & CPU-only     & 2950s           & 1228s    & 125 GFLOP/s  & 30\%             \\
with 1x NVIDIA\regtd V100 (PCI-E)                       & 1 GPU  & 1790s           & 68s     & 2271 GFLOP/s & 32\%             \\
with 2x NVIDIA\regtd V100 (PCI-E)                       & 2 GPU  & 1770s           & 48s     & 3185 GFLOP/s & 22\%             \\
\midrule
Intel\regtd Xeon\tradem E5-2660 v3 , 2.4 GHz, 20 Cores  & CPU-only     & 1601s           & 614s    & 250 GFLOP/s  & 30\%             \\
with 1x NVIDIA\regtd V100 (PCI-E)                       & 1 GPU  & 1086s           & 100s     & 1516 GFLOP/s & 22\%             \\
with 2x NVIDIA\regtd V100 (PCI-E)                       & 2 GPU  & 1017s           & 30s     & 5188 GFLOP/s & 37\%             \\
\midrule
Intel\regtd Xeon\tradem Phi 7210 , 1.3 GHz, 64 Cores &     & 1774s                  & 334s     & 459 GFLOPS/s & 17\%                 \\
\midrule
\textbf{One Piz Daint Node} & & & & & \\
Intel\regtd	Xeon\regtd E5-2690v3 , 2.6GHz, 12 Cores & CPU-only & 2415s & 980s & 157 GFLOP/s & 31\% \\
with 1x NVIDIA\regtd P100 (PCI-E)                       & 1 GPU  & 1592s           & 158s     & 973 GFLOP/s & 21\%             \\
\bottomrule
\end{tabular}


  \caption{FMM kernel node-level performance on various platforms. On platforms with GPUs
    we compare the performance with and without GPUs. The theoretical
    peak performance used for calculating the fraction of peak
    performance corresponds to the utilized device.}
  \label{tab:node_level_performance}
  \vspace{-2em}
\end{table*}

In the following, we will take a closer look at the performance of the
FMM kernels, discussed in Sect.~\ref{sec:fmm} and~\ref{sec:gpu}, on both GPUs and different CPU platforms. We will first explain how we made measurements and then discuss the results.

\subsubsection{Measuring the Node-Level Performance}
Measuring the node-level results for the FMM solver alone presents several challenges.
Instead of a few large kernels, we are executing millions of small FMM kernels overall.
Additionally, one FMM kernel alone will never utilize the complete device. On the CPU, each FMM kernel is executed by just one core. We cannot assume that the other cores will always be busy executing an FMM kernel as well. On the GPU, one kernel will utilize only up to 8 Streaming Multiprocessors (SM). The NVIDIA P100 GPU contains 56 of these SMs, each of which is analogue to a SIMD-enabled processor core.

In order to see how well we utilize the given hardware with the FMM kernels, 
we focus not on the performance of a single kernel. We rather focus on
the overall performance while computing the gravity during the
GPU-accelerated FMM part of the code.

In order to calculate both the GFLOP/s and the fraction of peak performance, we need to know the number of floating point operations executed while calculating the gravity, as well as the time required to do so.
The first piece of information is easy to collect. Each FMM kernel always executes a constant number of floating point operations. We count the number of kernel launches in each HPX thread and accumulate this number until the end of the simulation. We can further record whether a kernel was executed on the CPU or the GPU.
Due to the interleaving of kernels and the general lack of synchronization points between the gravity solver and the fluid solver, the amount of runtime spent in the FMM solver is more difficult to obtain.
To measure it, we run the simulation multiple times;
first, on the CPU without any GPUs. We collect profiling data with perf to get an estimation of the fraction of the runtime spent within the FMM kernels and thus the gravity solver. With this information we calculate the fraction of the 
runtime spent outside the gravity solver.
Afterwards we repeat the run -- without perf -- and multiply its total runtime with the earlier obtained runtime fractions to get both the time spent in the gravity solver and the time spent in other methods.
With this information, as well as the counters for the FMM kernel launches, we can now calculate the GFLOP/s achieved by the CPU when executing the FMM kernels. To get the same information for the GPUs, we include them in a third run of the same simulation.
Using the GPUs, only the runtime of the gravity solver will improve since the rest of the code does not benefit from them. Thus, by subtracting the runtime spent outside of the FMM kernels in the CPU-only run from the total runtime of the third run, we can estimate the overall runtime of the GPU-enabled FMM kernels
and with that the GFLOP/s we achieve overall during their execution.

For all results in this work, we employ the same V1309 scenario and
double precision calculations. The level 14 octree discretization considered here
will serve as the baseline for scaling runs.


\subsubsection{Results}
The results of our node-level runs can be found in Tab.~\ref{tab:node_level_performance}.
Switching to a stencil-based approach for the FMM instead of the old
interaction-lists, the fraction of time spent in the two main FMM
kernels shrank considerably. On the Intel Xeon E5-2660 v3 with 20
cores, they now only make up 38\% of the total runtime. On the Intel
Xeon Phi 7210 this difference is even higher, with the FMM only making
up 20\% of the total runtime. This is most likely due to the fact that
the other less optimized parts of \octo make fewer use of the SIMD capabilites that the Xeon Phi offers and are thus running a lot slower. This reduces the overall fraction of the FMM runtime compared to the rest of the code.

Nevertheless, we achieve a significant fraction of peak performance on all devices.
On the CPU-side, the Xeon Phi 7210 achieves the most GFLOP/s within the FMM kernels. Since it lowers its frequency to 1.1 GHz during AVX-intensive calculations,
the real achieved fraction of peak performance may be significantly higher than 17\%. 
We have assumed the base (unthrottled) clock rate shown in the table for calculating the theoretical peak performance of the CPU devices.
Other than running a specific Vc version that supports AVX512 on Xeon
Phi, we did not adapt the code. However, we attain a reasonable
fraction of peak performance on this difficult hardware. 
On the AVX2 CPUs we reach about 30\%.

We tested GPU performance of the FMM kernels in multiple hardware configurations; we used either 10 or 20 cores in combination with either one or two V100 GPUs.
Using two V100 GPUs, an insufficient number of cores affects
performance. With 20 cores and two GPUs we achieve 37\% of the
combined V100 peak performance. Reducing to 10 cores, the performance drops to 22\% of the peak. Then, the GPUs get starved of work, since the 10 cores have a lot of tasks to work on and cannot launch enough kernels on the GPU.

Simultaneously, when utilizing one V100 GPU managed by 10 cores, we
achieve 32\% of peak performance on the GPU. But using one V100 with
20 CPU cores, the performance decreases, achieving only 22\% peak: The
number of threads used to fill the CUDA streams of the GPU directly affects the
performance.
This effect can be explained by the way we handle CUDA streams. Each CPU thread manages a certain number of CUDA streams. When launching a kernel, a thread first checks whether all of the CUDA streams it manages are busy. If not, the kernel will be launched on the GPU using an idle stream. Otherwise, the kernel will be executed on the CPU by the current CPU worker thread. Executing an FMM kernel on the CPU takes significantly longer than on the GPU, as one CPU kernel will be executed on one core. In a CPU-only setting all cores are working on FMM kernels of different octree nodes.

With 20 cores and one V100, the CPU threads first fill all 128 streams with 128 kernel launches. Launching the next kernels, the GPU has not finished yet, and the CPU threads start to work on FMM kernels themselves. This leads to starvation of the GPU for a short period of time, as the CPU threads are not launching more work on the GPU in the meantime. Having two V100 offsets the problem, as the cores are less likely to work on the FMM themselves: It is more likely that there is a free CUDA stream available. 
We analyzed the number of kernels launched on the GPU to provide further data on this. Using 20 cores and one V100 we launch 97.4995\% of all multipole-multipole FMM kernels on the GPU. Using 10 cores and one V100 this number increases to 99.9997\%. Considering that a CPU FMM execution on one core takes longer than on the GPU and that during this time no other GPU kernels are launched in the meantime, the small difference in percentage can cause a large performance impact. 
This is a current limitation of our implementation and will be addressed in the next version of Octo-Tiger: There is no reason not to launch multiple FMM kernels in one stream if there is no empty stream available. This would lead to 100\% of the FMM kernels launched on the GPU independent of the CPU hardware.

Since Piz Daint is our target system, we also evaluated performance on one of its nodes, using 128 CUDA streams. 
For comparison, 99.5207\% of all multipole-multipole FMM kernels were launched on the GPU. We achieve about 21\% of peak performance on the GPU. In summary, we were able to demonstrate that the uncommon approach of launching many small kernels is a valid way to utilize the GPU.
\vspace{-0.5em}
\subsection{Scaling results}
\label{sec:results}
All of the presented distributed scaling results were obtained on Piz Daint 
at the Swiss National Supercomputing Centre.
Table~\ref{tab:hardware} lists the hardware configuration of Piz Daint.

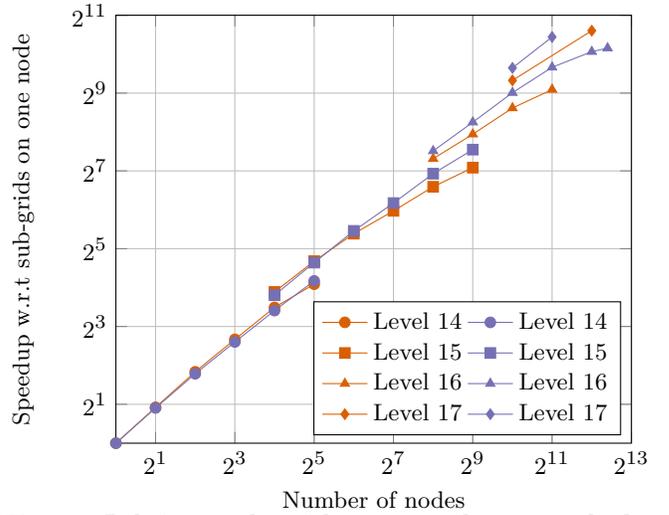
\begin{figure}[tbp]
\begin{tikzpicture}
\begin{axis}[
  ylabel={Speedup w.r.t sub-grids on one node},
  xlabel={Number of nodes},
  legend style={at={(0.980,0.175)},anchor=east},    
  legend columns=2,
  xmode=log,
  ymode=log,
  log basis x={2},
  log basis y={2},
  xmajorgrids=true,
  ymajorgrids=true,
  xmin=1,
  xmax=8192,
  ymin=1,
  ymax=2048
  ]
  \addplot[cbred,mark=*]          table [x=Nodes, y=speedup, col sep=comma] {speedup_daint_level_14.csv};
  \addplot[cbblue,mark=*]         table [x=Nodes, y=speedup, col sep=comma] {speedup_daint_level_14_libfabric.csv};
  \addplot[cbred,mark=square*]    table [x=Nodes, y=speedup, col sep=comma] {speedup_daint_level_15.csv};
  \addplot[cbblue,mark=square*]   table [x=Nodes, y=speedup, col sep=comma] {speedup_daint_level_15_libfabric.csv};
  \addplot[cbred,mark=triangle*]  table [x=Nodes, y=speedup, col sep=comma] {speedup_daint_level_16.csv};
  \addplot[cbblue,mark=triangle*] table [x=Nodes, y=speedup, col sep=comma] {speedup_daint_level_16_libfabric.csv};
  \addplot[cbred,mark=diamond*]   table [x=Nodes, y=speedup, col sep=comma] {speedup_daint_level_17.csv};
  \addplot[cbblue,mark=diamond*]  table [x=Nodes, y=speedup, col sep=comma] {speedup_daint_level_17_libfabric.csv};
  \legend{Level $14$\\Level $14$\\Level $15$\\Level $15$\\Level $16$\\Level $16$\\Level $17$\\Level $17$\\};
\end{axis}
\end{tikzpicture}

    \vspace{-1.5em}
    \caption{Relative speedup with respect to the processed sub-grids on one node for level 14. The \textcolor{cbred}{red lines} show the results using HPX's MPI parcelport and the \textcolor{cbblue}{blue lines} using HPX's libfabric parcelport, respectively. Note that for level 16 and level 17 some data points are missing due to restricted node hours for development projects.}
    \label{fig:speedup_daint}
\end{figure}

\begin{table}[tb]
\begin{tabular}{l|c}
\toprule
    & Piz Daint                                                                                \\
\midrule
CPU & $1$ $\times$ Intel\regtd Xeon\tradem E$5$-$2690$ v3, 2.60GHz, 12 cores   \\
GPU & $1$ $\times$ NVIDIA\regtd Tesla\regtd P$100$                                               \\
RAM & $64$ GB                     \\
IC  & Cray Aries routing and communications ASIC                    \\
\bottomrule
\end{tabular}
  \caption{Configuration of Piz Daint.}
  \label{tab:hardware}
  \vspace{-2em}
\end{table}
For the scalability analysis of \octo different levels of refinement of the V1309 scenario were run, as shown in Tab.~\ref{tab:level_of_refinement}. 
A level $13$ restart file, which takes less than an hour to generate on an Intel(R) Xeon(R) Gold 5118 CPU @ 2.30GHz, was used as the basis for all runs. For all levels the restart file for level $13$ was read and refined to higher levels of resolution through conservative interpolation of the evolved variables. The number of nodes was increased in powers of two $(1,2,4,\ldots)$ up to $4096$ nodes with a maximum of
$5400$ which corresponds to the full system on Piz Daint. All runs utilized 12 CPU cores on each node, i.e. up to $64,800$ cores for the full-system run.
The simulations started at level $14$, the smallest that fits on a single Piz Daint node with respect to memory while still consisting of an acceptable number of sub-grids to expose sufficient parallelism. The number of nodes was increased by a power of two until the scaling saturated due to too little work per node. 
Higher refinement levels were then run on the largest overlapping node counts
to produce the graph shown in Fig.~\ref{fig:speedup_daint}, where the speedup 
is calculated with respect to the number of processed sub-grids per 
second on one node at level $14$. 
The graph therefore shows a combination of weak scaling as the level of
refinement increases and strong scaling for each refinement level as the node count
increases. 
Weak scaling is clearly very good, with close to optimal improvements with 
successive refinement levels. 
Strong scaling tails off as the amount of sub-grids for each level becomes too small to generate sufficient work for all CPUs/GPUs. 

\begin{table}[tb]
\begin{filecontents*}{levels.csv}
levels,sub-grids, memory usage (GB)
13,5417,8.001
14,10928,16.365
15,42947,56.921
16,223953, 271.938
17,1500457, 2305.921
\end{filecontents*}
\pgfplotstabletypeset[
col sep = comma,
columns/B/.style={dec sep align={c|}, column type/.add={}{|}},
columns/C/.style={dec sep align={c|}, column type/.add={}{|}},
every head row/.style={before row=\toprule,after row=\midrule},
every last row/.style={after row=\bottomrule},
display columns/0/.style={string type,column name={Level of refinement}, column type={c|}},
display columns/1/.style={column type={c|}},
display columns/2/.style={column type={c}}
]
{levels.csv}
  \caption{Number of tree nodes (sub-grids) per level of refinement (LoR) and the memory usage of the corresponding level.}
  \label{tab:level_of_refinement}
  \vspace{-2em}
\end{table}

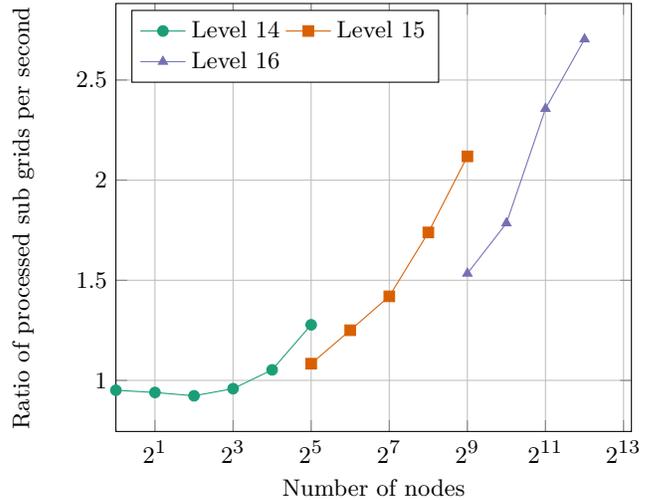
\begin{figure}[tb]
\begin{tikzpicture}
\begin{axis}[
  ylabel={Ratio of processed sub grids per second},
  xlabel={Number of nodes},
  legend style={at={(0.03,0.9)},anchor=west},
  legend columns=2,
  xmode=log,
  xmajorgrids=true,
  ymajorgrids=true,
  log basis x=2,
  xmin = 1
  ]
  \addplot[cbgreen,mark=*] table [x, y, col sep=comma] {difference_level_14.csv};
  \addplot[cbred,mark=square*] table [x, y, col sep=comma] {difference_level_15.csv};
  \addplot[cbblue,mark=triangle*] table [x, y, col sep=comma] {difference_level_16.csv};
  \legend{Level $14$\\Level $15$\\Level $16$\\};
\end{axis}
\end{tikzpicture}

  \vspace{-0.5em}
  \caption{\label{fig:speedup_subgrids_single_node}Ratio of processed sub grids per second between HPXs libfabric and MPI Parcelport on Piz Daint
  (higher numbers mean libfabric is faster).
  }
  \vspace{-1em}
\end{figure}

\vspace{-0.5em}
\subsection{Network performance results}
Figure~\ref{fig:speedup_daint} shows the speedup of both libfabric and
MPI parcelport on Piz Daint. The libfabric parcelport scales much
better than the MPI parcelport and in fact outperforms it by a factor
of almost 3 for the largest runs.
At level 17 on 1024 nodes, the libfabric version achieves a (weak) scalability
of 78.4\% of the efficiency of the reference value of level 14 on 1 node;
for 2048 nodes the value drops to 68.1\%. Where there is enough work to keep processors busy and overlap communication for large runs, impressive scaling can be observed.
At level 16 the efficiency values range from
71.4\% at 256 nodes down to 21.2\% on 5400 nodes where the communication dominates.
The performance difference between the number of sub-grids processed per second
for the two parcelports increases with higher node counts and refinement level,
a sure sign that communication is responsible for causing delays that
prevent the processing cores from getting work done. Each
increase in the refinement level can, due to AMR, increase the total number of grids by up to a
factor of 8; see Tab.~\ref{tab:level_of_refinement} for
the actual values. This causes a near quadratic increase in the total
number of halos exchanged.
As the node count increases, the probability of a halo exchange
increases linearly, and it
is therefore no surprise that reduced communication latency leads to
the large gains observed.
The improvement in communication is due to all of the following changes:
\begin{compactitem}
\item Explicit use of RMA for the transfer of halo buffers.
\item Lower latency on send and receive of all parcels and execution of RMA transfers.
\item Direct control of all memory copies for send/receive buffers between
the HPX runtime and the libfabric driver.
\item Reduced overhead between receipt of a transfer/message completion event and
subsequent setting of a \textit{ready} future.
\item Thread-safe lock-free interface between the HPX scheduling loop
and the libfabric API with polling for network progress/completions
integrated into the HPX task scheduling loop.
\end{compactitem}
\smallskip

It is important to note that the timing results shown are for the
core calculation steps that exchange halos, and the figures do not
include regridding steps or I/O that also make heavy use of communication. Including them would further illustrate the effectiveness
of the networking layer:
Start-up timings of the main solver at refinement level 16 and 17
were in fact reduced by an order of magnitude using the
libfabric parcelport, increasing the efficiency of refining the initial restart file of level 13 to the desired level of resolution.
Note further that some data points at level 16 and 17 for large runs are missing as the start-up time
consumed the limited node hours available to their execution.

The communication speedups shown have not separately considered
the effects of thread pools and the scheduling of network progress
on the rates of injection or the handling of messages.
When running on Piz Daint with 12 worker threads executing tasks, any thread might need
to send data across the network. In general, the injection of data into
send queues does not cause problems unless many threads are attempting
to do so concurrently and the send queues are full.
The receipt of data, however, must be performed by polling of
completion queues. This can only take place in-between the execution of other tasks.
Thus, if all cores are busy with work, no polling is done,
and if no work is available, all cores compete for access to the network.
The effects can be observed in Fig.~\ref{fig:speedup_subgrids_single_node}
where the libfabric parcelport causes a slight reduction in
performance for lower node counts.
With GPUs doing most of the work, CPU cores can be reserved for network
processing, and the job of polling can be restricted to a subset of cores
that have no other (longer running) tasks to execute.
HPX supports partitioning of a compute node into separate thread pools with different
responsibilities; the effects of this will be investigated further to
see whether reducing contention between cores helps to restore the lost performance.



\vspace{-0.5em}
\section{Conclusions and Future Work}

As the core contributions of this paper, we have demonstrated
node-level and distributed performance of \octo, an astrophysics code
simulating a stellar binary merger. We have shown excellent scaling up
to the full system on Piz Daint and improved network performance based
on the libfabric library. The high-level abstractions we employ, in
particular HPX and Vc, demonstrate how portability in heterogeneous
HPC systems is possible. This is the first time an HPX application was run on a full system of a GPU-accelerated supercomputer. This work also has several implications for parallel 
programming for future architectures. The asynchronous many-task 
runtime systems like HPX are a powerful, viable, and promising addition to 
the current landscape of parallel programming models. We show that it is 
not only possible to utilize these emerging tools to perform on the largest 
scales, but also that it might even be desirable to leverage the latency 
hiding, finer-grained parallelism and natural support for heterogeneity 
that the asynchronous many-task model exposes.

In particular, we have significantly increased node-level
performance of the originally most compute hungry part of \octo, the
gravitational solver. Our optimizations have demonstrated excellent
node-level performance on different HPC compute nodes
with heterogeneous hardware, including multi-GPU systems and KNL. We
have achieved up to 37\% of the peak performance on two NVIDIA V100 GPUs, and 17\% of peak on a KNL system. To achieve high node-level
performance for the full simulation, we will also port the remaining
part, the hydrodynamics solver, to GPUs.

The distributed scaling results have been obtained
within a development project on Piz Daint and thus with severely 
limited compute time. The excellent results presented in
this paper have already built the foundation for a production proposal
that will enable us to target full-resolution simulations with
impact on physics.

Despite the significant performance improvement replacing MPI with
libfabric, there are more networking improvements under development that have not been incorporated 
into \octo yet. This includes the use of user-controlled RMA buffers that allow the user to 
instruct the runtime that certain memory regions will be used repeatedly
for communication (and thus amortize memory pinning/registration costs).
Integration of such features into the channel abstraction may prove
to reduce latencies further and is an area we will explore.

With respect to the astrophysical application, we have already
developed  a radiation transport module for \octo\ based 
on the two moment approach adapted by \cite{skinner}. This will be
required to simulate the V1309 merger with high accuracy. What remains is to fully debug 
and verify this module and to port the implementation to GPUs.

Finally,  our full-scale
simulations will be able to predict the outcome of mergers that have
not yet happened: These simulations will useful for comparison with
future ``red nova'' contact-binary merger events.  Two contact-binary
systems have been suggested as future mergers, KIC 9832227
\cite{molnar,socia} and TY Pup \cite{sarotsakulchai}. Other candidate
systems will be discovered with the new all-sky surveys such as the
Zwicky Transient Facility (ZTF) and the Large Synoptic Survey
Telescope (LSST).




\begin{acks}
We thank the Swiss National Supercomputing Centre and the National Energy
Research Scientific Computing Center for providing is with the node hours to
run the simulations and the Center of Computation \& Technology at Louisiana
State University for supporting this work. Portions of this research was
conducted with high performance computational resources provided by the
Louisiana Optical Network Infrastructure (http://www.loni.org).
\end{acks}

\bibliographystyle{ACM-Reference-Format} 
\bibliography{references} 


\begin{thebibliography}{62}


\ifx \showCODEN    \undefined \def \showCODEN     #1{\unskip}     \fi
\ifx \showDOI      \undefined \def \showDOI       #1{#1}\fi
\ifx \showISBNx    \undefined \def \showISBNx     #1{\unskip}     \fi
\ifx \showISBNxiii \undefined \def \showISBNxiii  #1{\unskip}     \fi
\ifx \showISSN     \undefined \def \showISSN      #1{\unskip}     \fi
\ifx \showLCCN     \undefined \def \showLCCN      #1{\unskip}     \fi
\ifx \shownote     \undefined \def \shownote      #1{#1}          \fi
\ifx \showarticletitle \undefined \def \showarticletitle #1{#1}   \fi
\ifx \showURL      \undefined \def \showURL       {\relax}        \fi
\providecommand\bibfield[2]{#2}
\providecommand\bibinfo[2]{#2}
\providecommand\natexlab[1]{#1}
\providecommand\showeprint[2][]{arXiv:#2}

\bibitem[\protect\citeauthoryear{??}{nan}{[n. d.]}]%
        {nandez_2008}
 \bibinfo{year}{[n. d.]}\natexlab{}.
\newblock \bibinfo{title}{Red Giant and Main Sequence Binary (V1309 Sco)}.
\newblock
  \bibinfo{howpublished}{\url{https://www.sharcnet.ca/~jnandez/simulations.html}}.
\newblock
\newblock
\shownote{Accessed: 2019-03-14.}


\bibitem[\protect\citeauthoryear{??}{lom}{[n. d.]}]%
        {lombardi_2016}
 \bibinfo{year}{[n. d.]}\natexlab{}.
\newblock \bibinfo{title}{StarSmasher - a Smoothed Particle Hydrodynamics
  code}.
\newblock
  \bibinfo{howpublished}{\url{https://jalombar.github.io/starsmasher/}}.
\newblock
\newblock
\shownote{Accessed: 2019-03-14.}


\bibitem[\protect\citeauthoryear{Agullo, Bramas, Coulaud, Darve, Messner, and
  Takahashi}{Agullo et~al\mbox{.}}{2016a}]%
        {agullo2016task}
\bibfield{author}{\bibinfo{person}{Emmanuel Agullo}, \bibinfo{person}{Berenger
  Bramas}, \bibinfo{person}{Olivier Coulaud}, \bibinfo{person}{Eric Darve},
  \bibinfo{person}{Matthias Messner}, {and} \bibinfo{person}{Toru Takahashi}.}
  \bibinfo{year}{2016}\natexlab{a}.
\newblock \showarticletitle{Task-based FMM for heterogeneous architectures}.
\newblock \bibinfo{journal}{\emph{Concurrency and Computation: Practice and
  Experience}} \bibinfo{volume}{28}, \bibinfo{number}{9}
  (\bibinfo{year}{2016}), \bibinfo{pages}{2608--2629}.
\newblock


\bibitem[\protect\citeauthoryear{Agullo, Bramas, Coulaud, Khannouz, and
  Stanisic}{Agullo et~al\mbox{.}}{2016b}]%
        {agullo2016taskreport}
\bibfield{author}{\bibinfo{person}{Emmanuel Agullo},
  \bibinfo{person}{B{\'e}renger Bramas}, \bibinfo{person}{Olivier Coulaud},
  \bibinfo{person}{Martin Khannouz}, {and} \bibinfo{person}{Luka Stanisic}.}
  \bibinfo{year}{2016}\natexlab{b}.
\newblock \emph{\bibinfo{title}{Task-based fast multipole method for clusters
  of multicore processors}}.
\newblock \bibinfo{thesistype}{Ph.D. Dissertation}. \bibinfo{school}{Inria
  Bordeaux Sud-Ouest}.
\newblock


\bibitem[\protect\citeauthoryear{Augonnet, Thibault, Namyst, and
  Wacrenier}{Augonnet et~al\mbox{.}}{2011}]%
        {augonnet2011starpu}
\bibfield{author}{\bibinfo{person}{C{\'e}dric Augonnet},
  \bibinfo{person}{Samuel Thibault}, \bibinfo{person}{Raymond Namyst}, {and}
  \bibinfo{person}{Pierre-Andr{\'e} Wacrenier}.}
  \bibinfo{year}{2011}\natexlab{}.
\newblock \showarticletitle{StarPU: a unified platform for task scheduling on
  heterogeneous multicore architectures}.
\newblock \bibinfo{journal}{\emph{Concurrency and Computation: Practice and
  Experience}} \bibinfo{volume}{23}, \bibinfo{number}{2}
  (\bibinfo{year}{2011}), \bibinfo{pages}{187--198}.
\newblock


\bibitem[\protect\citeauthoryear{Bauer, Treichler, Slaughter, and Aiken}{Bauer
  et~al\mbox{.}}{2012}]%
        {bauer2012legion}
\bibfield{author}{\bibinfo{person}{Michael Bauer}, \bibinfo{person}{Sean
  Treichler}, \bibinfo{person}{Elliott Slaughter}, {and} \bibinfo{person}{Alex
  Aiken}.} \bibinfo{year}{2012}\natexlab{}.
\newblock \showarticletitle{Legion: Expressing locality and independence with
  logical regions}. In \bibinfo{booktitle}{\emph{SC'12: Proceedings of the
  International Conference on High Performance Computing, Networking, Storage
  and Analysis}}. IEEE, \bibinfo{pages}{1--11}.
\newblock


\bibitem[\protect\citeauthoryear{Biddiscombe, Heller, Bikineev, and
  Kaiser}{Biddiscombe et~al\mbox{.}}{2017}]%
        {ac:2017}
\bibfield{author}{\bibinfo{person}{John Biddiscombe}, \bibinfo{person}{Thomas
  Heller}, \bibinfo{person}{Anton Bikineev}, {and} \bibinfo{person}{Hartmut
  Kaiser}.} \bibinfo{year}{2017}\natexlab{}.
\newblock \showarticletitle{{Zero Copy Serialization using RMA in the
  Distributed Task-Based HPX runtime}}. In \bibinfo{booktitle}{\emph{14th
  International Conference on Applied Computing}}. \bibinfo{publisher}{IADIS,
  International Association for Development of the Information Society}.
\newblock


\bibitem[\protect\citeauthoryear{Blumofe, Joerg, Kuszmaul, Leiserson, Randall,
  and Zhou}{Blumofe et~al\mbox{.}}{1996}]%
        {blumofe1996cilk}
\bibfield{author}{\bibinfo{person}{Robert~D Blumofe},
  \bibinfo{person}{Christopher~F Joerg}, \bibinfo{person}{Bradley~C Kuszmaul},
  \bibinfo{person}{Charles~E Leiserson}, \bibinfo{person}{Keith~H Randall},
  {and} \bibinfo{person}{Yuli Zhou}.} \bibinfo{year}{1996}\natexlab{}.
\newblock \showarticletitle{Cilk: An efficient multithreaded runtime system}.
\newblock \bibinfo{journal}{\emph{Journal of parallel and distributed
  computing}} \bibinfo{volume}{37}, \bibinfo{number}{1} (\bibinfo{year}{1996}),
  \bibinfo{pages}{55--69}.
\newblock


\bibitem[\protect\citeauthoryear{Bosilca, Bouteiller, Danalis, Faverge,
  H{\'e}rault, and Dongarra}{Bosilca et~al\mbox{.}}{2013}]%
        {bosilca2013parsec}
\bibfield{author}{\bibinfo{person}{George Bosilca}, \bibinfo{person}{Aurelien
  Bouteiller}, \bibinfo{person}{Anthony Danalis}, \bibinfo{person}{Mathieu
  Faverge}, \bibinfo{person}{Thomas H{\'e}rault}, {and} \bibinfo{person}{Jack~J
  Dongarra}.} \bibinfo{year}{2013}\natexlab{}.
\newblock \showarticletitle{Parsec: Exploiting heterogeneity to enhance
  scalability}.
\newblock \bibinfo{journal}{\emph{Computing in Science \& Engineering}}
  \bibinfo{volume}{15}, \bibinfo{number}{6} (\bibinfo{year}{2013}),
  \bibinfo{pages}{36--45}.
\newblock


\bibitem[\protect\citeauthoryear{Bryan, Norman, O'Shea, Abel, Wise, Turk,
  Reynolds, Collins, Wang, Skillman, et~al\mbox{.}}{Bryan
  et~al\mbox{.}}{2014}]%
        {bryan2014enzo}
\bibfield{author}{\bibinfo{person}{Greg~L Bryan}, \bibinfo{person}{Michael~L
  Norman}, \bibinfo{person}{Brian~W O'Shea}, \bibinfo{person}{Tom Abel},
  \bibinfo{person}{John~H Wise}, \bibinfo{person}{Matthew~J Turk},
  \bibinfo{person}{Daniel~R Reynolds}, \bibinfo{person}{David~C Collins},
  \bibinfo{person}{Peng Wang}, \bibinfo{person}{Samuel~W Skillman},
  {et~al\mbox{.}}} \bibinfo{year}{2014}\natexlab{}.
\newblock \showarticletitle{Enzo: An adaptive mesh refinement code for
  astrophysics}.
\newblock \bibinfo{journal}{\emph{The Astrophysical Journal Supplement Series}}
  \bibinfo{volume}{211}, \bibinfo{number}{2} (\bibinfo{year}{2014}),
  \bibinfo{pages}{19}.
\newblock


\bibitem[\protect\citeauthoryear{Chamberlain, Callahan, and Zima}{Chamberlain
  et~al\mbox{.}}{2007}]%
        {chamberlain2007parallel}
\bibfield{author}{\bibinfo{person}{Bradford~L Chamberlain},
  \bibinfo{person}{David Callahan}, {and} \bibinfo{person}{Hans~P Zima}.}
  \bibinfo{year}{2007}\natexlab{}.
\newblock \showarticletitle{Parallel programmability and the chapel language}.
\newblock \bibinfo{journal}{\emph{The International Journal of High Performance
  Computing Applications}} \bibinfo{volume}{21}, \bibinfo{number}{3}
  (\bibinfo{year}{2007}), \bibinfo{pages}{291--312}.
\newblock


\bibitem[\protect\citeauthoryear{Choi, Chandramowlishwaran, Madduri, and
  Vuduc}{Choi et~al\mbox{.}}{2014}]%
        {choi2014cpu}
\bibfield{author}{\bibinfo{person}{Jee Choi}, \bibinfo{person}{Aparna
  Chandramowlishwaran}, \bibinfo{person}{Kamesh Madduri}, {and}
  \bibinfo{person}{Richard Vuduc}.} \bibinfo{year}{2014}\natexlab{}.
\newblock \showarticletitle{A cpu: Gpu hybrid implementation and model-driven
  scheduling of the fast multipole method}. In
  \bibinfo{booktitle}{\emph{Proceedings of Workshop on General Purpose
  Processing Using GPUs}}. ACM, \bibinfo{pages}{64}.
\newblock


\bibitem[\protect\citeauthoryear{{Colella} and {Woodward}}{{Colella} and
  {Woodward}}{1984}]%
        {1984JCoPh..54..174C}
\bibfield{author}{\bibinfo{person}{P. {Colella}} {and} \bibinfo{person}{P.~R.
  {Woodward}}.} \bibinfo{year}{1984}\natexlab{}.
\newblock \showarticletitle{{The Piecewise Parabolic Method (PPM) for
  Gas-Dynamical Simulations}}.
\newblock \bibinfo{journal}{\emph{J. Comput. Phys.}}  \bibinfo{volume}{54}
  (\bibinfo{date}{Sept.} \bibinfo{year}{1984}), \bibinfo{pages}{174--201}.
\newblock
\urldef\tempurl%
\url{https://doi.org/10.1016/0021-9991(84)90143-8}
\showDOI{\tempurl}


\bibitem[\protect\citeauthoryear{Dagum and Menon}{Dagum and Menon}{1998}]%
        {Dagum:1998:OIA:615255.615542}
\bibfield{author}{\bibinfo{person}{Leonardo Dagum} {and}
  \bibinfo{person}{Ramesh Menon}.} \bibinfo{year}{1998}\natexlab{}.
\newblock \showarticletitle{OpenMP: An Industry-Standard API for Shared-Memory
  Programming}.
\newblock \bibinfo{journal}{\emph{IEEE Comput. Sci. Eng.}} \bibinfo{volume}{5},
  \bibinfo{number}{1} (\bibinfo{date}{Jan.} \bibinfo{year}{1998}),
  \bibinfo{pages}{46--55}.
\newblock
\showISSN{1070-9924}
\urldef\tempurl%
\url{https://doi.org/10.1109/99.660313}
\showDOI{\tempurl}


\bibitem[\protect\citeauthoryear{Dai\ss}{Dai\ss}{2018}]%
        {master_thesis_2018_daiss}
\bibfield{author}{\bibinfo{person}{Gregor Dai\ss}.}
  \bibinfo{year}{2018}\natexlab{}.
\newblock \emph{\bibinfo{title}{Octo-Tiger: Binary Star Systems with HPX on
  Nvidia P100}}.
\newblock Master thesis. \bibinfo{school}{Universität Stuttgart}.
\newblock


\bibitem[\protect\citeauthoryear{Dan, Rosswog, Guillochon, and
  Ramirez-Ruiz}{Dan et~al\mbox{.}}{2011}]%
        {2011apj...737...89d}
\bibfield{author}{\bibinfo{person}{Marius Dan}, \bibinfo{person}{Stephan
  Rosswog}, \bibinfo{person}{James Guillochon}, {and} \bibinfo{person}{Enrico
  Ramirez-Ruiz}.} \bibinfo{year}{2011}\natexlab{}.
\newblock \showarticletitle{{Prelude to A Double Degenerate Merger: The Onset
  of Mass Transfer and Its Impact on Gravitational Waves and Surface
  Detonations}}.
\newblock \bibinfo{journal}{\emph{Astrophysical Journal (ApJ)}}
  \bibinfo{volume}{737}, \bibinfo{number}{2, art. id 89}
  (\bibinfo{year}{2011}).
\newblock
\showISSN{0004-637X}
\urldef\tempurl%
\url{https://doi.org/10.1088/0004-637X/737/2/89}
\showDOI{\tempurl}
\newblock
\shownote{\url{http://adsabs.harvard.edu/abs/2011ApJ...737...89D}.}


\bibitem[\protect\citeauthoryear{de~Supinski Michael~Klemm}{de~Supinski
  Michael~Klemm}{2017}]%
        {openmptr}
\bibfield{author}{\bibinfo{person}{Bronis~R. de Supinski Michael~Klemm}.}
  \bibinfo{year}{2017}\natexlab{}.
\newblock \bibinfo{booktitle}{\emph{OpenMP Technical Report 6:Version 5.0
  Preview 2}}.
\newblock \bibinfo{type}{{T}echnical {R}eport}. \bibinfo{institution}{OpenMP
  Architecture Review Board}.
\newblock


\bibitem[\protect\citeauthoryear{Desprésa and Labourasse}{Desprésa and
  Labourasse}{2015}]%
        {despres2015}
\bibfield{author}{\bibinfo{person}{Bruno Desprésa} {and}
  \bibinfo{person}{Emmanuel Labourasse}.} \bibinfo{year}{2015}\natexlab{}.
\newblock \showarticletitle{{Angular Momentum Preserving Cell-Centered
  Lagrangian and Eulerian Schemes on Arbitrary Grids}}.
\newblock \bibinfo{journal}{\emph{J. Comput. Phys.}}  \bibinfo{volume}{290}
  (\bibinfo{year}{2015}), \bibinfo{pages}{28--54}.
\newblock
\showISSN{0021-9991}
\urldef\tempurl%
\url{https://doi.org/10.1016/j.jcp.2015.02.032}
\showDOI{\tempurl}
\newblock
\shownote{\url{https://dx.doi.org/10.1016/j.jcp.2015.02.032}.}


\bibitem[\protect\citeauthoryear{Edwards, Trott, and Sunderland}{Edwards
  et~al\mbox{.}}{2014}]%
        {CarterEdwards20143202}
\bibfield{author}{\bibinfo{person}{H.~Carter Edwards},
  \bibinfo{person}{Christian~R. Trott}, {and} \bibinfo{person}{Daniel
  Sunderland}.} \bibinfo{year}{2014}\natexlab{}.
\newblock \showarticletitle{Kokkos: Enabling manycore performance portability
  through polymorphic memory access patterns}.
\newblock \bibinfo{journal}{\emph{J. Parallel and Distrib. Comput.}}
  \bibinfo{volume}{74}, \bibinfo{number}{12} (\bibinfo{year}{2014}),
  \bibinfo{pages}{3202 -- 3216}.
\newblock
\showISSN{0743-7315}
\urldef\tempurl%
\url{https://doi.org/10.1016/j.jpdc.2014.07.003}
\showDOI{\tempurl}
\newblock
\shownote{Domain-Specific Languages and High-Level Frameworks for
  High-Performance Computing.}


\bibitem[\protect\citeauthoryear{{Even} and {Tohline}}{{Even} and
  {Tohline}}{2009}]%
        {even_and_tohline}
\bibfield{author}{\bibinfo{person}{Wesley {Even}} {and}
  \bibinfo{person}{Joel~E. {Tohline}}.} \bibinfo{year}{2009}\natexlab{}.
\newblock \showarticletitle{{Constructing Synchronously Rotating Double White
  Dwarf Binaries}}.
\newblock \bibinfo{journal}{\emph{The Astrophysical Journal Supplement Series}}
   \bibinfo{volume}{184} (\bibinfo{date}{Oct} \bibinfo{year}{2009}),
  \bibinfo{pages}{248--263}.
\newblock
\urldef\tempurl%
\url{https://doi.org/10.1088/0067-0049/184/2/248}
\showDOI{\tempurl}
\showeprint[arxiv]{astro-ph.SR/0908.2116}


\bibitem[\protect\citeauthoryear{Faber, Lombardi, and Rasio}{Faber
  et~al\mbox{.}}{2010}]%
        {faber2010starcrash}
\bibfield{author}{\bibinfo{person}{Joshua Faber}, \bibinfo{person}{Jamie
  Lombardi}, {and} \bibinfo{person}{Fred Rasio}.}
  \bibinfo{year}{2010}\natexlab{}.
\newblock \showarticletitle{StarCrash: 3-d Evolution of Self-gravitating Fluid
  Systems}.
\newblock \bibinfo{journal}{\emph{Astrophysics Source Code Library}}
  (\bibinfo{year}{2010}).
\newblock


\bibitem[\protect\citeauthoryear{Germain, McCorquodale, Parker, and
  Johnson}{Germain et~al\mbox{.}}{2000}]%
        {germain2000uintah}
\bibfield{author}{\bibinfo{person}{J~Davison de~St Germain},
  \bibinfo{person}{John McCorquodale}, \bibinfo{person}{Steven~G Parker}, {and}
  \bibinfo{person}{Christopher~R Johnson}.} \bibinfo{year}{2000}\natexlab{}.
\newblock \showarticletitle{Uintah: A massively parallel problem solving
  environment}. In \bibinfo{booktitle}{\emph{Proceedings the Ninth
  International Symposium on High-Performance Distributed Computing}}. IEEE,
  \bibinfo{pages}{33--41}.
\newblock


\bibitem[\protect\citeauthoryear{{Hachisu}}{{Hachisu}}{1986}]%
        {hachisu}
\bibfield{author}{\bibinfo{person}{Izumi {Hachisu}}.}
  \bibinfo{year}{1986}\natexlab{}.
\newblock \showarticletitle{{A Versatile Method for Obtaining Structures of
  Rapidly Rotating Stars. II. Three-dimensional Self-consistent Field Method}}.
\newblock \bibinfo{journal}{\emph{The Astrophysical Journal Supplement Series}}
   \bibinfo{volume}{62} (\bibinfo{date}{Nov} \bibinfo{year}{1986}),
  \bibinfo{pages}{461}.
\newblock
\urldef\tempurl%
\url{https://doi.org/10.1086/191148}
\showDOI{\tempurl}


\bibitem[\protect\citeauthoryear{Heller, Kaiser, Diehl, Fey, and
  Schweitzer}{Heller et~al\mbox{.}}{2016}]%
        {heller:hpc_2016}
\bibfield{author}{\bibinfo{person}{Thomas Heller}, \bibinfo{person}{Hartmut
  Kaiser}, \bibinfo{person}{Patrick Diehl}, \bibinfo{person}{Dietmar Fey},
  {and} \bibinfo{person}{Marc~Alexander Schweitzer}.}
  \bibinfo{year}{2016}\natexlab{}.
\newblock \showarticletitle{Closing the Performance Gap with Modern C++}. In
  \bibinfo{booktitle}{\emph{High Performance Computing}}
  \emph{(\bibinfo{series}{Lecture Notes in Computer Science})},
  \bibfield{editor}{\bibinfo{person}{Michaela Taufer}, \bibinfo{person}{Bernd
  Mohr}, {and} \bibinfo{person}{Julian~M. Kunkel}} (Eds.),
  Vol.~\bibinfo{volume}{9945}. \bibinfo{publisher}{Springer International
  Publishing}, \bibinfo{pages}{18--31}.
\newblock


\bibitem[\protect\citeauthoryear{Heller, Kaiser, and Iglberger}{Heller
  et~al\mbox{.}}{2012}]%
        {heller:2012}
\bibfield{author}{\bibinfo{person}{Thomas Heller}, \bibinfo{person}{Hartmut
  Kaiser}, {and} \bibinfo{person}{Klaus Iglberger}.}
  \bibinfo{year}{2012}\natexlab{}.
\newblock \showarticletitle{{Application of the {ParalleX} Execution Model to
  Stencil-Based Problems}}.
\newblock \bibinfo{journal}{\emph{Computer Science - Research and Development}}
  \bibinfo{volume}{28}, \bibinfo{number}{2-3} (\bibinfo{year}{2012}),
  \bibinfo{pages}{253--261}.
\newblock
\showISSN{1865-2034}
\urldef\tempurl%
\url{https://doi.org/10.1007/s00450-012-0217-1}
\showDOI{\tempurl}
\newblock
\shownote{\url{https://stellar.cct.lsu.edu/pubs/isc2012.pdf}.}


\bibitem[\protect\citeauthoryear{Heller, Kaiser, Sch\"{a}fer, and Fey}{Heller
  et~al\mbox{.}}{2013}]%
        {Heller:2013:UHL:2530268.2530269}
\bibfield{author}{\bibinfo{person}{Thomas Heller}, \bibinfo{person}{Hartmut
  Kaiser}, \bibinfo{person}{Andreas Sch\"{a}fer}, {and}
  \bibinfo{person}{Dietmar Fey}.} \bibinfo{year}{2013}\natexlab{}.
\newblock \showarticletitle{{Using HPX and LibGeoDecomp for Scaling HPC
  Applications on Heterogeneous Supercomputers}}. In
  \bibinfo{booktitle}{\emph{Proceedings of the ACM/IEEE Workshop on Latest
  Advances in Scalable Algorithms for Large-Scale Systems (ScalA, SC
  Workshop)}} \emph{(\bibinfo{series}{art. id 1})}.
\newblock
\showISBNx{978-1-4503-2508-0}
\urldef\tempurl%
\url{https://doi.org/10.1145/2530268.2530269}
\showDOI{\tempurl}
\newblock
\shownote{\url{https://stellar.cct.lsu.edu/pubs/scala13.pdf}.}


\bibitem[\protect\citeauthoryear{Heller, Lelbach, Huck, Biddiscombe, Grubel,
  Koniges, Kretz, Marcello, Pfander, Serio, Frank, Clayton, Pflüger, Eder, and
  Kaiser}{Heller et~al\mbox{.}}{2019}]%
        {heller_gb}
\bibfield{author}{\bibinfo{person}{Thomas Heller},
  \bibinfo{person}{Bryce~Adelstein Lelbach}, \bibinfo{person}{Kevin~A Huck},
  \bibinfo{person}{John Biddiscombe}, \bibinfo{person}{Patricia Grubel},
  \bibinfo{person}{Alice~E Koniges}, \bibinfo{person}{Matthias Kretz},
  \bibinfo{person}{Dominic Marcello}, \bibinfo{person}{David Pfander},
  \bibinfo{person}{Adrian Serio}, \bibinfo{person}{Juhan Frank},
  \bibinfo{person}{Geoffrey~C Clayton}, \bibinfo{person}{Dirk Pflüger},
  \bibinfo{person}{David Eder}, {and} \bibinfo{person}{Hartmut Kaiser}.}
  \bibinfo{year}{2019}\natexlab{}.
\newblock \showarticletitle{Harnessing billions of tasks for a scalable
  portable hydrodynamic simulation of the merger of two stars}.
\newblock \bibinfo{journal}{\emph{The International Journal of High Performance
  Computing Applications}} (\bibinfo{year}{2019}).
\newblock
\urldef\tempurl%
\url{https://doi.org/10.1177/1094342018819744}
\showDOI{\tempurl}
\showeprint{https://doi.org/10.1177/1094342018819744}
\newblock
\shownote{published online.}


\bibitem[\protect\citeauthoryear{Kaiser, Heller, Bourgeois, and Fey}{Kaiser
  et~al\mbox{.}}{2015}]%
        {espm2_2015}
\bibfield{author}{\bibinfo{person}{Hartmut Kaiser}, \bibinfo{person}{Thomas
  Heller}, \bibinfo{person}{Daniel Bourgeois}, {and} \bibinfo{person}{Dietmar
  Fey}.} \bibinfo{year}{2015}\natexlab{}.
\newblock \showarticletitle{Higher-level Parallelization for Local and
  Distributed Asynchronous Task-Based Programming}. In
  \bibinfo{booktitle}{\emph{First International Workshop on Extreme Scale
  Programming Models and Middleware}}. \bibinfo{pages}{29--37}.
\newblock
\showISBNx{978-1-4503-3996-4}
\urldef\tempurl%
\url{https://doi.org/10.1145/2832241.2832244}
\showDOI{\tempurl}
\newblock
\shownote{\url{https://stellar.cct.lsu.edu/pubs/executors_espm2_2015.pdf}.}


\bibitem[\protect\citeauthoryear{Kaiser, Heller, Lelbach, Serio, and
  Fey}{Kaiser et~al\mbox{.}}{2014}]%
        {hpx_pgas_2014}
\bibfield{author}{\bibinfo{person}{Hartmut Kaiser}, \bibinfo{person}{Thomas
  Heller}, \bibinfo{person}{Bryce~Adelstein Lelbach}, \bibinfo{person}{Adrian
  Serio}, {and} \bibinfo{person}{Dietmar Fey}.}
  \bibinfo{year}{2014}\natexlab{}.
\newblock \showarticletitle{{HPX: A Task Based Programming Model in a Global
  Address Space}}. In \bibinfo{booktitle}{\emph{Proceedings of the
  International Conference on Partitioned Global Address Space Programming
  Models (PGAS)}} \emph{(\bibinfo{series}{art. id 6})}.
\newblock
\showISBNx{978-1-4503-3247-7}
\urldef\tempurl%
\url{https://doi.org/10.1145/2676870.2676883}
\showDOI{\tempurl}
\newblock
\shownote{\url{https://stellar.cct.lsu.edu/pubs/pgas14.pdf}.}


\bibitem[\protect\citeauthoryear{Kale and Krishnan}{Kale and Krishnan}{1993}]%
        {kale1993charm++}
\bibfield{author}{\bibinfo{person}{Laxmikant~V Kale} {and}
  \bibinfo{person}{Sanjeev Krishnan}.} \bibinfo{year}{1993}\natexlab{}.
\newblock \showarticletitle{CHARM++: a portable concurrent object oriented
  system based on C++}. In \bibinfo{booktitle}{\emph{OOPSLA}},
  Vol.~\bibinfo{volume}{93}. Citeseer, \bibinfo{pages}{91--108}.
\newblock


\bibitem[\protect\citeauthoryear{Kretz}{Kretz}{2015}]%
        {k2015}
\bibfield{author}{\bibinfo{person}{Matthias Kretz}.}
  \bibinfo{year}{2015}\natexlab{}.
\newblock \emph{\bibinfo{title}{{Extending C++ for Explicit Data-Parallel
  Programming via SIMD Vector Types}}}.
\newblock \bibinfo{thesistype}{Ph.D. Dissertation}. \bibinfo{school}{Goethe
  University Frankfurt}.
\newblock
\urldef\tempurl%
\url{https://doi.org/10.13140/RG.2.1.2355.4323}
\showDOI{\tempurl}
\newblock
\shownote{\url{http://publikationen.ub.uni-frankfurt.de/frontdoor/index/index/docId/38415}.}


\bibitem[\protect\citeauthoryear{Kurganov and Tadmor}{Kurganov and
  Tadmor}{2000}]%
        {2000jcoph.160..241k}
\bibfield{author}{\bibinfo{person}{Alexander Kurganov} {and}
  \bibinfo{person}{Eitan Tadmor}.} \bibinfo{year}{2000}\natexlab{}.
\newblock \showarticletitle{{New High-Resolution Central Schemes for Nonlinear
  Conservation Laws and Convection-Diffusion Equations}}.
\newblock \bibinfo{journal}{\emph{J. Comput. Phys.}} \bibinfo{volume}{160},
  \bibinfo{number}{1} (\bibinfo{year}{2000}), \bibinfo{pages}{241--282}.
\newblock
\showISSN{0021-9991}
\urldef\tempurl%
\url{https://doi.org/10.1006/jcph.2000.6459}
\showDOI{\tempurl}
\newblock
\shownote{\url{https://dx.doi.org/10.1006/jcph.2000.6459}.}


\bibitem[\protect\citeauthoryear{Ltaief and Yokota}{Ltaief and Yokota}{2014}]%
        {ltaief2014data}
\bibfield{author}{\bibinfo{person}{Hatem Ltaief} {and} \bibinfo{person}{Rio
  Yokota}.} \bibinfo{year}{2014}\natexlab{}.
\newblock \showarticletitle{Data-driven execution of fast multipole methods}.
\newblock \bibinfo{journal}{\emph{Concurrency and Computation: Practice and
  Experience}} \bibinfo{volume}{26}, \bibinfo{number}{11}
  (\bibinfo{year}{2014}), \bibinfo{pages}{1935--1946}.
\newblock


\bibitem[\protect\citeauthoryear{MacLeod, Ostriker, and Stone}{MacLeod
  et~al\mbox{.}}{2018a}]%
        {macleod_2018_2}
\bibfield{author}{\bibinfo{person}{Morgan MacLeod}, \bibinfo{person}{Eve~C.
  Ostriker}, {and} \bibinfo{person}{James~M. Stone}.}
  \bibinfo{year}{2018}\natexlab{a}.
\newblock \showarticletitle{Bound Outflows, Unbound Ejecta, and the Shaping of
  Bipolar Remnants during Stellar Coalescence}.
\newblock \bibinfo{journal}{\emph{The Astrophysical Journal}}
  \bibinfo{volume}{868}, \bibinfo{number}{2} (\bibinfo{date}{dec}
  \bibinfo{year}{2018}), \bibinfo{pages}{136}.
\newblock
\urldef\tempurl%
\url{https://doi.org/10.3847/1538-4357/aae9eb}
\showDOI{\tempurl}


\bibitem[\protect\citeauthoryear{MacLeod, Ostriker, and Stone}{MacLeod
  et~al\mbox{.}}{2018b}]%
        {macleod_2018}
\bibfield{author}{\bibinfo{person}{Morgan MacLeod}, \bibinfo{person}{Eve~C.
  Ostriker}, {and} \bibinfo{person}{James~M. Stone}.}
  \bibinfo{year}{2018}\natexlab{b}.
\newblock \showarticletitle{Runaway Coalescence at the Onset of Common Envelope
  Episodes}.
\newblock \bibinfo{journal}{\emph{The Astrophysical Journal}}
  \bibinfo{volume}{863}, \bibinfo{number}{1} (\bibinfo{date}{aug}
  \bibinfo{year}{2018}), \bibinfo{pages}{5}.
\newblock
\urldef\tempurl%
\url{https://doi.org/10.3847/1538-4357/aacf08}
\showDOI{\tempurl}


\bibitem[\protect\citeauthoryear{{Marcello}}{{Marcello}}{2017}]%
        {octotiger_fmm}
\bibfield{author}{\bibinfo{person}{D.~C. {Marcello}}.}
  \bibinfo{year}{2017}\natexlab{}.
\newblock \showarticletitle{{A Very Fast and Angular Momentum Conserving Tree
  Code}}.
\newblock \bibinfo{journal}{\emph{Astronomical Journal}}
  \bibinfo{volume}{154}, Article \bibinfo{articleno}{92} (\bibinfo{date}{Sept.}
  \bibinfo{year}{2017}), \bibinfo{numpages}{92}~pages.
\newblock
\urldef\tempurl%
\url{https://doi.org/10.3847/1538-3881/aa7b2f}
\showDOI{\tempurl}
\showeprint[arxiv]{astro-ph.IM/1706.06989}


\bibitem[\protect\citeauthoryear{Marcello, Kadam, Clayton, Frank, Kaiser, and
  Motl}{Marcello et~al\mbox{.}}{2016}]%
        {octotiger_apcs_2016}
\bibfield{author}{\bibinfo{person}{Dominic~C. Marcello},
  \bibinfo{person}{Kundan Kadam}, \bibinfo{person}{Geoffrey~C. Clayton},
  \bibinfo{person}{Juhan Frank}, \bibinfo{person}{Hartmut Kaiser}, {and}
  \bibinfo{person}{Patrick~M. Motl}.} \bibinfo{year}{2016}\natexlab{}.
\newblock \showarticletitle{{Introducing Octo-tiger/HPX: Simulating Interacting
  Binaries with Adaptive Mesh Refinement and the Fast Multipole Method}}. In
  \bibinfo{booktitle}{\emph{Proceedings of the International Conference on
  Accretion Processes in Cosmic Sources}}.
\newblock
\newblock
\shownote{\url{http://apcs2016.iaps.inaf.it}.}


\bibitem[\protect\citeauthoryear{{Marcello} and {Tohline}}{{Marcello} and
  {Tohline}}{2012}]%
        {marcello_and_tohline}
\bibfield{author}{\bibinfo{person}{Dominic~C. {Marcello}} {and}
  \bibinfo{person}{Joel~E. {Tohline}}.} \bibinfo{year}{2012}\natexlab{}.
\newblock \showarticletitle{{A Numerical Method for Studying Super-Eddington
  Mass Transfer in Double White Dwarf Binaries}}.
\newblock \bibinfo{journal}{\emph{The Astrophysical Journal Supplement Series}}
   \bibinfo{volume}{199}, Article \bibinfo{articleno}{35} (\bibinfo{date}{Apr}
  \bibinfo{year}{2012}), \bibinfo{numpages}{35}~pages.
\newblock
\urldef\tempurl%
\url{https://doi.org/10.1088/0067-0049/199/2/35}
\showDOI{\tempurl}
\showeprint[arxiv]{astro-ph.SR/1404.6208}


\bibitem[\protect\citeauthoryear{{Mason, E.}, {Diaz, M.}, {Williams, R. E.},
  {Preston, G.}, and {Bensby, T.}}{{Mason, E.} et~al\mbox{.}}{2010}]%
        {mason_2010}
\bibfield{author}{\bibinfo{person}{{Mason, E.}}, \bibinfo{person}{{Diaz, M.}},
  \bibinfo{person}{{Williams, R. E.}}, \bibinfo{person}{{Preston, G.}}, {and}
  \bibinfo{person}{{Bensby, T.}}} \bibinfo{year}{2010}\natexlab{}.
\newblock \showarticletitle{{The peculiar nova V1309 Scorpii/nova Scorpii 2008*
  - A candidate twin of V838 Monocerotis}}.
\newblock \bibinfo{journal}{\emph{A\&A}}  \bibinfo{volume}{516}
  (\bibinfo{year}{2010}), \bibinfo{pages}{A108}.
\newblock
\urldef\tempurl%
\url{https://doi.org/10.1051/0004-6361/200913610}
\showDOI{\tempurl}


\bibitem[\protect\citeauthoryear{{Molnar}, {Van Noord}, {Kinemuchi},
  {Smolinski}, {Alexander}, {Cook}, {Jang}, {Kobulnicky}, {Spedden}, and
  {Steenwyk}}{{Molnar} et~al\mbox{.}}{2017}]%
        {molnar}
\bibfield{author}{\bibinfo{person}{L.~A. {Molnar}}, \bibinfo{person}{D.~M. {Van
  Noord}}, \bibinfo{person}{K. {Kinemuchi}}, \bibinfo{person}{J.~P.
  {Smolinski}}, \bibinfo{person}{C.~E. {Alexander}}, \bibinfo{person}{E.~M.
  {Cook}}, \bibinfo{person}{B. {Jang}}, \bibinfo{person}{H.~A. {Kobulnicky}},
  \bibinfo{person}{C.~J. {Spedden}}, {and} \bibinfo{person}{S.~D. {Steenwyk}}.}
  \bibinfo{year}{2017}\natexlab{}.
\newblock \showarticletitle{{Prediction of a Red Nova Outburst in KIC
  9832227}}.
\newblock \bibinfo{journal}{\emph{Astrophysical Journal}}
  \bibinfo{volume}{840}, Article \bibinfo{articleno}{1} (\bibinfo{date}{May}
  \bibinfo{year}{2017}).
\newblock
\urldef\tempurl%
\url{https://doi.org/10.3847/1538-4357/aa6ba7}
\showDOI{\tempurl}
\showeprint[arxiv]{astro-ph.SR/1704.05502}


\bibitem[\protect\citeauthoryear{Motl, Tohline, and Frank}{Motl
  et~al\mbox{.}}{2002}]%
        {Motl_2002}
\bibfield{author}{\bibinfo{person}{Patrick~M. Motl}, \bibinfo{person}{Joel~E.
  Tohline}, {and} \bibinfo{person}{Juhan Frank}.}
  \bibinfo{year}{2002}\natexlab{}.
\newblock \showarticletitle{Numerical Methods for the Simulation of Dynamical
  Mass Transfer in Binaries}.
\newblock \bibinfo{journal}{\emph{The Astrophysical Journal Supplement Series}}
  \bibinfo{volume}{138}, \bibinfo{number}{1} (\bibinfo{date}{jan}
  \bibinfo{year}{2002}), \bibinfo{pages}{121--148}.
\newblock
\urldef\tempurl%
\url{https://doi.org/10.1086/324159}
\showDOI{\tempurl}


\bibitem[\protect\citeauthoryear{Orr, Beckmann, Reinhardt, and Wood}{Orr
  et~al\mbox{.}}{2014}]%
        {orr2014fine}
\bibfield{author}{\bibinfo{person}{Marc~S Orr}, \bibinfo{person}{Bradford~M
  Beckmann}, \bibinfo{person}{Steven~K Reinhardt}, {and}
  \bibinfo{person}{David~A Wood}.} \bibinfo{year}{2014}\natexlab{}.
\newblock \showarticletitle{Fine-grain task aggregation and coordination on
  GPUs}.
\newblock \bibinfo{journal}{\emph{ACM SIGARCH Computer Architecture News}}
  \bibinfo{volume}{42}, \bibinfo{number}{3} (\bibinfo{year}{2014}),
  \bibinfo{pages}{181--192}.
\newblock


\bibitem[\protect\citeauthoryear{Pejcha, Metzger, and Tomida}{Pejcha
  et~al\mbox{.}}{2015}]%
        {pejcha2015cool}
\bibfield{author}{\bibinfo{person}{Ond{\v{r}}ej Pejcha},
  \bibinfo{person}{Brian~D Metzger}, {and} \bibinfo{person}{Kengo Tomida}.}
  \bibinfo{year}{2015}\natexlab{}.
\newblock \showarticletitle{Cool and luminous transients from mass-losing
  binary stars}.
\newblock \bibinfo{journal}{\emph{Monthly Notices of the Royal Astronomical
  Society}} \bibinfo{volume}{455}, \bibinfo{number}{4} (\bibinfo{year}{2015}),
  \bibinfo{pages}{4351--4372}.
\newblock


\bibitem[\protect\citeauthoryear{Pejcha, Metzger, Tyles, and Tomida}{Pejcha
  et~al\mbox{.}}{2017}]%
        {pejcha_2017}
\bibfield{author}{\bibinfo{person}{Ondrej Pejcha}, \bibinfo{person}{Brian~D.
  Metzger}, \bibinfo{person}{Jacob~G. Tyles}, {and} \bibinfo{person}{Kengo
  Tomida}.} \bibinfo{year}{2017}\natexlab{}.
\newblock \showarticletitle{Pre-explosion Spiral Mass Loss of a Binary Star
  Merger}.
\newblock \bibinfo{journal}{\emph{The Astrophysical Journal}}
  \bibinfo{volume}{850}, \bibinfo{number}{1} (\bibinfo{date}{nov}
  \bibinfo{year}{2017}), \bibinfo{pages}{59}.
\newblock
\urldef\tempurl%
\url{https://doi.org/10.3847/1538-4357/aa95b9}
\showDOI{\tempurl}


\bibitem[\protect\citeauthoryear{Pfander, Dai\ss, Marcello, Kaiser, and
  Pfl\"{u}ger}{Pfander et~al\mbox{.}}{2018}]%
        {Pfander18accelerating}
\bibfield{author}{\bibinfo{person}{David Pfander}, \bibinfo{person}{Gregor
  Dai\ss}, \bibinfo{person}{Dominic Marcello}, \bibinfo{person}{Hartmut
  Kaiser}, {and} \bibinfo{person}{Dirk Pfl\"{u}ger}.}
  \bibinfo{year}{2018}\natexlab{}.
\newblock \showarticletitle{Accelerating {Octo-Tiger}: Stellar Mergers on
  {Intel Knights Landing} with {HPX}}. In \bibinfo{booktitle}{\emph{Proceedings
  of the International Workshop on OpenCL}} \emph{(\bibinfo{series}{IWOCL
  '18})}. \bibinfo{publisher}{ACM}, \bibinfo{address}{New York, NY, USA},
  Article \bibinfo{articleno}{19}, \bibinfo{numpages}{8}~pages.
\newblock
\showISBNx{978-1-4503-6439-3}
\urldef\tempurl%
\url{https://doi.org/10.1145/3204919.3204938}
\showDOI{\tempurl}


\bibitem[\protect\citeauthoryear{Pritchard, Harvey, Choi, Swaro, and
  Tiffany}{Pritchard et~al\mbox{.}}{2016}]%
        {pritchard2016gni}
\bibfield{author}{\bibinfo{person}{Howard Pritchard}, \bibinfo{person}{Evan
  Harvey}, \bibinfo{person}{Sung-Eun Choi}, \bibinfo{person}{James Swaro},
  {and} \bibinfo{person}{Zachary Tiffany}.} \bibinfo{year}{2016}\natexlab{}.
\newblock \showarticletitle{The GNI provider layer for OFI libfabric}. In
  \bibinfo{booktitle}{\emph{Proceedings of Cray User Group Meeting, CUG}},
  Vol.~\bibinfo{volume}{2016}.
\newblock


\bibitem[\protect\citeauthoryear{{Sarotsakulchai}, {Qian}, {Soonthornthum},
  {Zhou}, {Zhang}, {Reichart}, {Haislip}, {Kouprianov}, and
  {Poshyachinda}}{{Sarotsakulchai} et~al\mbox{.}}{2018}]%
        {sarotsakulchai}
\bibfield{author}{\bibinfo{person}{T. {Sarotsakulchai}}, \bibinfo{person}{S.-B.
  {Qian}}, \bibinfo{person}{B. {Soonthornthum}}, \bibinfo{person}{X. {Zhou}},
  \bibinfo{person}{J. {Zhang}}, \bibinfo{person}{D.~E. {Reichart}},
  \bibinfo{person}{J.~B. {Haislip}}, \bibinfo{person}{V.~V. {Kouprianov}},
  {and} \bibinfo{person}{S. {Poshyachinda}}.} \bibinfo{year}{2018}\natexlab{}.
\newblock \showarticletitle{{TY Pup: A Low-mass-ratio and Deep Contact Binary
  as a Progenitor Candidate of Luminous Red Novae}}.
\newblock \bibinfo{journal}{\emph{Journal of Astrophysics}}
  \bibinfo{volume}{156}, Article \bibinfo{articleno}{199} (\bibinfo{date}{Nov.}
  \bibinfo{year}{2018}), \bibinfo{numpages}{199}~pages.
\newblock
\urldef\tempurl%
\url{https://doi.org/10.3847/1538-3881/aadcfa}
\showDOI{\tempurl}
\showeprint[arxiv]{astro-ph.SR/1807.00478}


\bibitem[\protect\citeauthoryear{{Skinner} and {Ostriker}}{{Skinner} and
  {Ostriker}}{2013}]%
        {skinner}
\bibfield{author}{\bibinfo{person}{M.~Aaron {Skinner}} {and}
  \bibinfo{person}{Eve~C. {Ostriker}}.} \bibinfo{year}{2013}\natexlab{}.
\newblock \showarticletitle{{A Two-moment Radiation Hydrodynamics Module in
  Athena Using a Time-explicit Godunov Method}}.
\newblock \bibinfo{journal}{\emph{The Astrophysical Journal Supplement Series}}
   \bibinfo{volume}{206}, Article \bibinfo{articleno}{21} (\bibinfo{date}{Jun}
  \bibinfo{year}{2013}), \bibinfo{numpages}{21}~pages.
\newblock
\urldef\tempurl%
\url{https://doi.org/10.1088/0067-0049/206/2/21}
\showDOI{\tempurl}
\showeprint[arxiv]{astro-ph.IM/1306.0010}


\bibitem[\protect\citeauthoryear{{Socia}, {Welsh}, {Short}, {Orosz}, {Angione},
  {Windmiller}, {Caldwell}, and {Batalha}}{{Socia} et~al\mbox{.}}{2018}]%
        {socia}
\bibfield{author}{\bibinfo{person}{Q.~J. {Socia}}, \bibinfo{person}{W.~F.
  {Welsh}}, \bibinfo{person}{D.~R. {Short}}, \bibinfo{person}{J.~A. {Orosz}},
  \bibinfo{person}{R.~J. {Angione}}, \bibinfo{person}{G. {Windmiller}},
  \bibinfo{person}{D.~A. {Caldwell}}, {and} \bibinfo{person}{N.~M. {Batalha}}.}
  \bibinfo{year}{2018}\natexlab{}.
\newblock \showarticletitle{{KIC 9832227: Using Vulcan Data to Negate the 2022
  Red Nova Merger Prediction}}.
\newblock \bibinfo{journal}{\emph{Astrophysical Journal Letters}}
  \bibinfo{volume}{864}, Article \bibinfo{articleno}{L32}
  (\bibinfo{date}{Sept.} \bibinfo{year}{2018}), \bibinfo{numpages}{L32}~pages.
\newblock
\urldef\tempurl%
\url{https://doi.org/10.3847/2041-8213/aadc0d}
\showDOI{\tempurl}
\showeprint[arxiv]{astro-ph.SR/1809.02771}


\bibitem[\protect\citeauthoryear{{St{\c{e}}pie{\'n}}}{{St{\c{e}}pie{\'n}}}{2011}]%
        {stepien_2011}
\bibfield{author}{\bibinfo{person}{K. {St{\c{e}}pie{\'n}}}.}
  \bibinfo{year}{2011}\natexlab{}.
\newblock \showarticletitle{{Evolution of the progenitor binary of V1309
  Scorpii before merger}}.
\newblock \bibinfo{journal}{\emph{A\&A}}  \bibinfo{volume}{531}, Article
  \bibinfo{articleno}{A18} (\bibinfo{date}{Jul} \bibinfo{year}{2011}),
  \bibinfo{numpages}{A18}~pages.
\newblock
\urldef\tempurl%
\url{https://doi.org/10.1051/0004-6361/201116689}
\showDOI{\tempurl}
\showeprint[arxiv]{astro-ph.SR/1105.2627}


\bibitem[\protect\citeauthoryear{{STE$||$AR Group}}{{STE$||$AR Group}}{2017a}]%
        {hpx_git}
\bibfield{author}{\bibinfo{person}{{STE$||$AR Group}}.}
  \bibinfo{year}{2017}\natexlab{a}.
\newblock \bibinfo{title}{{HPX GitHub repository}}.
\newblock \bibinfo{howpublished}{\url{https://github.com/STEllAR-GROUP/hpx}}.
\newblock
\newblock
\shownote{Available under the Boost Software License 1.0 (a BSD-style open
  source license).}


\bibitem[\protect\citeauthoryear{{STE$||$AR Group}}{{STE$||$AR Group}}{2017b}]%
        {octotiger_git}
\bibfield{author}{\bibinfo{person}{{STE$||$AR Group}}.}
  \bibinfo{year}{2017}\natexlab{b}.
\newblock \bibinfo{title}{{OctoTiger AMR Framework GitHub repository}}.
\newblock
  \bibinfo{howpublished}{\url{https://github.com/STEllAR-GROUP/octotiger}}.
\newblock
\newblock
\shownote{Available under the Boost Software License 1.0 (a BSD-style open
  source license).}


\bibitem[\protect\citeauthoryear{Stone, Gardiner, Teuben, Hawley, and
  Simon}{Stone et~al\mbox{.}}{2008}]%
        {stone2008athena}
\bibfield{author}{\bibinfo{person}{James~M Stone}, \bibinfo{person}{Thomas~A
  Gardiner}, \bibinfo{person}{Peter Teuben}, \bibinfo{person}{John~F Hawley},
  {and} \bibinfo{person}{Jacob~B Simon}.} \bibinfo{year}{2008}\natexlab{}.
\newblock \showarticletitle{Athena: a new code for astrophysical MHD}.
\newblock \bibinfo{journal}{\emph{The Astrophysical Journal Supplement Series}}
  \bibinfo{volume}{178}, \bibinfo{number}{1} (\bibinfo{year}{2008}),
  \bibinfo{pages}{137}.
\newblock


\bibitem[\protect\citeauthoryear{{Stone, James M. and Gardiner, Thomas A. and
  Teuben, Peter}}{{Stone, James M. and Gardiner, Thomas A. and Teuben,
  Peter}}{2000}]%
        {athena++}
\bibfield{author}{\bibinfo{person}{{Stone, James M. and Gardiner, Thomas A. and
  Teuben, Peter}}.} \bibinfo{year}{2000}\natexlab{}.
\newblock \bibinfo{title}{{Athena++ radiation GRMHD code}}.
\newblock
  \bibinfo{howpublished}{\url{https://princetonuniversity.github.io/Athena-Cversion/}}.
\newblock
\newblock
\shownote{Available under the BSD 3-Clause "New" or "Revised" License.}


\bibitem[\protect\citeauthoryear{{Stone, James M. and Tomida, Kengo and White,
  Christopher and Felker, Kyle Gerard}}{{Stone, James M. and Tomida, Kengo and
  White, Christopher and Felker, Kyle Gerard}}{2016}]%
        {athena_repo}
\bibfield{author}{\bibinfo{person}{{Stone, James M. and Tomida, Kengo and
  White, Christopher and Felker, Kyle Gerard}}.}
  \bibinfo{year}{2016}\natexlab{}.
\newblock \bibinfo{title}{{Athena++ radiation GRMHD code}}.
\newblock
  \bibinfo{howpublished}{\url{http://princetonuniversity.github.io/athena/}}.
\newblock
\newblock
\shownote{Available under the BSD 3-Clause "New" or "Revised" License.}


\bibitem[\protect\citeauthoryear{{Tasker}, {Brunino}, {Mitchell}, {Michielsen},
  {Hopton}, {Pearce}, {Bryan}, and {Theuns}}{{Tasker} et~al\mbox{.}}{2008}]%
        {tasker}
\bibfield{author}{\bibinfo{person}{Elizabeth~J. {Tasker}},
  \bibinfo{person}{Riccardo {Brunino}}, \bibinfo{person}{Nigel~L. {Mitchell}},
  \bibinfo{person}{Dolf {Michielsen}}, \bibinfo{person}{Stephen {Hopton}},
  \bibinfo{person}{Frazer~R. {Pearce}}, \bibinfo{person}{Greg~L. {Bryan}},
  {and} \bibinfo{person}{Tom {Theuns}}.} \bibinfo{year}{2008}\natexlab{}.
\newblock \showarticletitle{{A test suite for quantitative comparison of
  hydrodynamic codes in astrophysics}}.
\newblock \bibinfo{journal}{\emph{Monthly Notices of the Royal Astronomical
  Society}} \bibinfo{volume}{390}, \bibinfo{number}{3} (\bibinfo{date}{Nov}
  \bibinfo{year}{2008}), \bibinfo{pages}{1267--1281}.
\newblock
\urldef\tempurl%
\url{https://doi.org/10.1111/j.1365-2966.2008.13836.x}
\showDOI{\tempurl}
\showeprint[arxiv]{astro-ph/0808.1844}


\bibitem[\protect\citeauthoryear{Thoman, Dichev, Heller, Iakymchuk, Aguilar,
  Hasanov, Gschwandtner, Lemarinier, Markidis, Jordan, et~al\mbox{.}}{Thoman
  et~al\mbox{.}}{2018}]%
        {thoman2018taxonomy}
\bibfield{author}{\bibinfo{person}{Peter Thoman}, \bibinfo{person}{Kiril
  Dichev}, \bibinfo{person}{Thomas Heller}, \bibinfo{person}{Roman Iakymchuk},
  \bibinfo{person}{Xavier Aguilar}, \bibinfo{person}{Khalid Hasanov},
  \bibinfo{person}{Philipp Gschwandtner}, \bibinfo{person}{Pierre Lemarinier},
  \bibinfo{person}{Stefano Markidis}, \bibinfo{person}{Herbert Jordan},
  {et~al\mbox{.}}} \bibinfo{year}{2018}\natexlab{}.
\newblock \showarticletitle{A taxonomy of task-based parallel programming
  technologies for high-performance computing}.
\newblock \bibinfo{journal}{\emph{The Journal of Supercomputing}}
  \bibinfo{volume}{74}, \bibinfo{number}{4} (\bibinfo{year}{2018}),
  \bibinfo{pages}{1422--1434}.
\newblock


\bibitem[\protect\citeauthoryear{{Tylenda}, {Hajduk}, {Kami{\'n}ski},
  {Udalski}, {Soszy{\'n}ski}, {Szyma{\'n}ski}, {Kubiak}, {Pietrzy{\'n}ski},
  {Poleski}, {Wyrzykowski}, and {Ulaczyk}}{{Tylenda} et~al\mbox{.}}{2011}]%
        {tylenda_2011}
\bibfield{author}{\bibinfo{person}{R. {Tylenda}}, \bibinfo{person}{M.
  {Hajduk}}, \bibinfo{person}{T. {Kami{\'n}ski}}, \bibinfo{person}{A.
  {Udalski}}, \bibinfo{person}{I. {Soszy{\'n}ski}}, \bibinfo{person}{M.~K.
  {Szyma{\'n}ski}}, \bibinfo{person}{M. {Kubiak}}, \bibinfo{person}{G.
  {Pietrzy{\'n}ski}}, \bibinfo{person}{R. {Poleski}}, \bibinfo{person}{{\L}.
  {Wyrzykowski}}, {and} \bibinfo{person}{K. {Ulaczyk}}.}
  \bibinfo{year}{2011}\natexlab{}.
\newblock \showarticletitle{{V1309 Scorpii: merger of a contact binary}}.
\newblock \bibinfo{journal}{\emph{A\&A}}  \bibinfo{volume}{528}, Article
  \bibinfo{articleno}{A114} (\bibinfo{date}{April} \bibinfo{year}{2011}),
  \bibinfo{numpages}{A114}~pages.
\newblock
\urldef\tempurl%
\url{https://doi.org/10.1051/0004-6361/201016221}
\showDOI{\tempurl}
\showeprint[arxiv]{astro-ph.SR/1012.0163}


\bibitem[\protect\citeauthoryear{Wang, Rubin, Sidelnik, and Yalamanchili}{Wang
  et~al\mbox{.}}{2016}]%
        {wang2016dynamic}
\bibfield{author}{\bibinfo{person}{Jin Wang}, \bibinfo{person}{Norm Rubin},
  \bibinfo{person}{Albert Sidelnik}, {and} \bibinfo{person}{Sudhakar
  Yalamanchili}.} \bibinfo{year}{2016}\natexlab{}.
\newblock \showarticletitle{Dynamic thread block launch: a lightweight
  execution mechanism to support irregular applications on GPUs}.
\newblock \bibinfo{journal}{\emph{ACM SIGARCH Computer Architecture News}}
  \bibinfo{volume}{43}, \bibinfo{number}{3} (\bibinfo{year}{2016}),
  \bibinfo{pages}{528--540}.
\newblock


\bibitem[\protect\citeauthoryear{YarKhan, Kurzak, and Dongarra}{YarKhan
  et~al\mbox{.}}{2011}]%
        {yarkhan2011quark}
\bibfield{author}{\bibinfo{person}{Asim YarKhan}, \bibinfo{person}{Jakub
  Kurzak}, {and} \bibinfo{person}{Jack Dongarra}.}
  \bibinfo{year}{2011}\natexlab{}.
\newblock \showarticletitle{Quark users’ guide: Queueing and runtime for
  kernels}.
\newblock \bibinfo{journal}{\emph{University of Tennessee Innovative Computing
  Laboratory Technical Report ICL-UT-11-02}} (\bibinfo{year}{2011}).
\newblock


\bibitem[\protect\citeauthoryear{Yokota, Barba, Narumi, and Yasuoka}{Yokota
  et~al\mbox{.}}{2013}]%
        {YOKOTA2013445}
\bibfield{author}{\bibinfo{person}{Rio Yokota}, \bibinfo{person}{L.A. Barba},
  \bibinfo{person}{Tetsu Narumi}, {and} \bibinfo{person}{Kenji Yasuoka}.}
  \bibinfo{year}{2013}\natexlab{}.
\newblock \showarticletitle{Petascale turbulence simulation using a highly
  parallel fast multipole method on GPUs}.
\newblock \bibinfo{journal}{\emph{Computer Physics Communications}}
  \bibinfo{volume}{184}, \bibinfo{number}{3} (\bibinfo{year}{2013}),
  \bibinfo{pages}{445 -- 455}.
\newblock
\showISSN{0010-4655}
\urldef\tempurl%
\url{https://doi.org/10.1016/j.cpc.2012.09.011}
\showDOI{\tempurl}


\bibitem[\protect\citeauthoryear{Zhang}{Zhang}{2014}]%
        {zhang2014asynchronous}
\bibfield{author}{\bibinfo{person}{Bo Zhang}.} \bibinfo{year}{2014}\natexlab{}.
\newblock \showarticletitle{Asynchronous task scheduling of the fast multipole
  method using various runtime systems}. In \bibinfo{booktitle}{\emph{2014
  Fourth Workshop on Data-Flow Execution Models for Extreme Scale Computing}}.
  IEEE, \bibinfo{pages}{9--16}.
\newblock


\end{thebibliography}

\onecolumn


\end{document}